\documentclass[preprint,floats,aps,12pt,superscriptaddress,nofootinbib,floatfix,iicol,sn-mathphys-num]{sn-jnl}
\usepackage{times} 
\usepackage{amssymb,amsmath}
\usepackage{csquotes}
\usepackage{array} %
\usepackage{multirow}%
\usepackage[usenames,dvipsnames]{color}
\usepackage{epstopdf}
\hypersetup{pdftitle={Interface-dominated sliding compound drops}, %
    pdfauthor={Dominik Thy and Jan Diekmann and Uwe Thiele},  %
pdfproducer={lateX},
pdfview=FitV,       %
pdfstartview=FitB,
linkcolor=blue,     %
citecolor=blue,     %
urlcolor=red,      %
breaklinks=true,    %
colorlinks=true,
citebordercolor=0 0 0,  %
filebordercolor=0 0 0,
linkbordercolor=0 0 0,
menubordercolor=0 0 0,
urlbordercolor=0 0 0,
pdfhighlight=/I,
pdfborder=0 0 0,   %
bookmarksopen=true,
bookmarksnumbered=true
}
\DeclareGraphicsExtensions{.jpg, .pdf, .tif, .png}
\usepackage[normalem]{ulem}
\usepackage{tcolorbox}

\renewcommand{\vec}[1]{\mathbf{#1}}

\newcommand{\tens}[1]{\mathbf{\underline{#1}}}

\usepackage{wrapfig}
\usepackage{mathtools}
\usepackage[section]{placeins}
\usepackage{xcolor}
\usepackage[separate-uncertainty=true]{siunitx}
\usepackage{tikz}
\usepackage{float}
\usepackage{tabularray}

\newcommand{\FG}{\mathcal{F}_\text{G}}
\newcommand{\hOT}{h_{12}}
\newcommand{\hTT}{h_{23}}
\newcommand{\gamOneTwo}{\ensuremath{\gamma_{12}}}
\newcommand{\gamOneThr}{\ensuremath{\gamma_{13}}}
\newcommand{\gamTwoThr}{\ensuremath{\gamma_{23}}}
\newcommand{\gamSOne}{\ensuremath{\gamma_{\text{s}1}}}
\newcommand{\gamSTwo}{\ensuremath{\gamma_{\text{s}2}}}
\newcommand{\gamSThr}{\ensuremath{\gamma_{\text{s}3}}}

\newcommand{\dFdhOT}{\frac{\delta \mathcal{F}}{\delta h_{12}}}
\newcommand{\dFdhTT}{\frac{\delta \mathcal{F}}{\delta h_{23}}}
\newcommand{\TildedFdhOT}{\frac{\delta \mathcal{F}}{\delta \Tilde{h}_{12}}}
\newcommand{\TildedFdhTT}{\frac{\delta \mathcal{F}}{\delta \Tilde{h}_{23}}}

\newcommand{\VolVer}{\ensuremath{\nu}}
\newcommand{\VisVer}{\ensuremath{\eta}}
\newcommand{\vel}{U} %

\begin{document}
\title{Interface-dominated sliding compound drops}

\author[1]{Dominik Thy}
\email{dominik.thy@uni-muenster.de}
\author[1]{Jan Diekmann}
\email{jan.diekmann@uni-muenster.de}
\author[1,2]{Uwe Thiele}
\email{u.thiele@uni-muenster.de}
\affil[1]{Institute of Theoretical Physics, University of M\"unster, Wilhelm-Klemm-Str.\ 9, 48149 M\"unster, Germany}
\affil[2]{Center for Data Science and Complexity (CDSC), University of M\"unster, Corrensstr.\ 2, 48149 M\"unster, Germany}

\abstract{ 
  We investigate compound drops composed of two immiscible nonvolatile partially wetting liquids that slide down an inclined homogeneous smooth solid substrate based on a mesoscopic hydrodynamic two-layer model in full-curvature formulation. {First, we briefly investigate the reference case of a liquid drop that stationarily slides on a liquid layer representing an adaptive substrate. Subsequently, we focus on sliding compound drops employing a parameter study to identify qualitative transitions between different drop configurations and time-periodic behavior as well as the underlying bifurcation structure. After analyzing the dependence of drop velocities and dynamic Young and Neumann angles on driving force, also the influence of the ratios of drop volumes and viscosities is  briefly considered. An accompanying discussion of the lateral dissipation profile is used to explain the encountered dependence of drop velocity on their configuration. Finally, we consider the time-periodic fusion-overtaking-splitting behavior found outside the existence range of the stationary sliding compound drops.}
\begin{center}
\begin{tcolorbox}[arc=0mm,colback=black!5!white,colframe=black!30!red,width=0.9\textwidth,title=Publication note]
This version of the article has been accepted for publication, after peer review but
is not the Version of Record and does not reflect all post-acceptance improvements and
corrections. The Version of Record is available online at\newline
\begin{center}
  Thy, Diekmann and Thiele,  \textit{The European Physical Journal E} 49, 72, 2026. \linebreak\textbf{doi}:  \href{https://doi.org/10.1140/epje/s10189-026-00612-9}{10.1140/epje/s10189-026-00612-9}.
\end{center}%
\end{tcolorbox}
\end{center}
}
\maketitle
\section{Introduction} \label{sec:intro}

When a drop of simple nonvolatile partially wetting liquid on a horizontal solid rigid substrate relaxes to thermodynamic equilibrium the drop shape converges to a spherical cap with equilibrium angles at the three-phase contact line that are related to the interface energies by Young's law \cite{Youn1805ptrs,Genn1985rmp}. It can be obtained as a horizontal force balance or via a minimization of the total interfacial free energy \cite{GennesBrochard-WyartQuere2004,Borm2009csaea}. However, tilting the assumed ideally uniform infinitely long substrate, reflexion symmetry along the substrate is broken implying that all such drops will slide at some speed driven by the gravitational downhill force. The velocity depends on drop size, substrate inclination, and the material parameters related to capillarity and wettability \cite{Genn1985rmp}. At least at small inclination, the drops will slide as stationary states, i.e., with constant shape and velocity. At larger velocity they might show oscillations or develop a cusp and undergo a pearling instability at their rear \cite{PoFL2001prl,LeDL2005jfm,SnAn2013arfm,EWGT2016prf,FBLD2019jcis,TsLT2023pf}. As a case of driven or forced wetting \cite{Thie2026epje} the stationary sliding drops feature dynamic contact angles that at the advancing and receding edge normally increase and decrease, respectively, with increasing velocity \cite{Genn1985rmp,KiLK2002jcis,SnAn2013arfm} as in different regimes captured by the cubic Cox-Voinov law 
\cite{Voin1976fd,Hock1983qjmam,Cox1986jfm,SnAn2013arfm,Moha2022jap} or a quadratic law  based on a variational energy-dissipation principle \cite{Pesc2018pf}. Recently, such studies focus on drops of simple liquids sliding on various types of adaptive substrates \cite{EDSD2021m}, e.g., polymer brushes of different thickness \cite{HDGT2024l,ZWLS2024am}, soft solid substrates \cite{KPLW2016pnasusa,AnSn2020arfm}, liquid-infused and liquid-like substrates \cite{GWXM2015l,SeMK2017sm,TKPS2017sm,KKCQ2017sm,LGGM2017pra,MJBW2025prl}, and dielectric substrates where charges are deposited \cite{LBSB2022np}.

Beside the drops of simple liquid on the various substrates there exist also many studies that consider free-surface films and drops that involve two immiscible liquids. For instance,  Refs.~\cite{BrMR1993l,PBMT2004pre,PBMT2005jcp,FiGo2005jcis,BaGS2005iecr,PBMT2006el,CrMa2009rmp,Ward2011pf,BCJP2013epje,YaKK2012prb, JPMW2014jem,XBRS2017sr,PBJS2018sr} consider two-layer films that may undergo dewetting. However, although after initial film rupture also the coarsening of small drops into larger ones and accompanying configuration changes are discussed, the final equilibrium shape of compound drops, i.e., symmetric and asymmetric drops that combine the two liquids are only mentioned in passing \cite{PBMT2005jcp,FiGo2005jcis}. Relatively well studied with closely related models are steady and spreading drops {of one liquid} on horizontal {substrates consisting of another (immiscible) liquid} \cite{Lued1869app,CrMa2006jcis,BCJP2013epje,JHKP2013sjam,HJKP2015jem} and sliding drops on inclined liquid substrates \cite{KrMi2003sjam}. Also the capillary leveling {of two-layer films is considered \cite{BLSR2021jfm}.}

Compound drops of two (and sometimes more) liquids at equilibrium are studied experimentally and with various theoretical models \cite{MaAP2002jfm,NTGD2012sm,BSNB2014l,ZCAG2016jcis,IDSS2017l,Kita2024jns,Nepo2021cocis,ZGLD2021jfm,YKDL2019sm}. In particular, Refs.~\cite{MaAP2002jfm,NTGD2012sm,ZCAG2016jcis,IDSS2017l} discuss different static drop configurations mainly based on macroscopic quantities, Ref.~\cite{BSNB2014l} uses a phase-field model,  Refs.~\cite{KrMi2003sjam,PBJS2018sr} employ piecewise thin-film descriptions for each fluid-fluid interface that are connected at the Neumann and Young points. However, then the equilibrium law is imposed, e.g., the static Neumann law at the three-phase contact points for a sliding drop  on a liquid substrate \cite{KrMi2003sjam}. A comparison to a mesoscopic hydrodynamic model and such a piecewise model is given in \cite{HJKP2015jem} for a slow relaxation dynamics.

Note that also more complex systems are considered, e.g., two-layer films of liquids with surfactants and/or evaporation \cite{DPAR1998ces,DPSA1998ces,PDAR1998ces} or involving viscoelastic and elastic materials \cite{MuSh2015sm}, two-layer films in nonisothermal situations \cite{PBMT2005jcp,NeSi2009prl,NeSi2017pf,NeSi2021prf,Nepo2021cocis}, films of liquid mixtures that phase-separate and dewet at the same time \cite{GeKr2003pps,GRBP2006pla,NaTh2010n,ThTL2013prl}. Such systems may show compound drops at late stages \cite{DGGR2021l,ATTG2024pre}.

{ 
  Besides the overdamped descriptions of mesoscopic hydrodynamics valid for interface-dominated dynamics at small Reynolds numbers discussed up to here, some influence of inertia can also be captured for such systems (see, e.g., section~VI.B of \cite{OrDB1997rmp} and \cite{CrMa2009rmp}). The simplest hydrodynamic model for a one-layer falling film that includes inertia is the Benney equation \cite{Benn1966jmp, SRTK2005jfm} that captures an inertia-driven surface-wave instability. More advanced modes are based on integrated layer models \cite{CrMa2009rmp}. Including inertia into descriptions of two-layer falling films, Refs.~\cite{CeSi2015ijonm, Klia1999jfm} study nonlinear surface waves and, in particular, Refs.~\cite{Yu2026wm, PoMa2023prf} focus on (solitary) rolling waves.
}

At equilibrium, the isothermal two-liquid system can in the macroscopic picture be completely characterized based on the six involved liquid-solid, liquid-liquid and liquid-gas interface energies with Laplace laws governing all fluid-fluid interface curvatures. All three-phase contact lines where three fluid-fluid interfaces meet are governed by Neumann's law that determines the angles between the interfaces, i.e., determines the absolute angles up to a solid-body rotation (cf.~chap.~6.1 and 6.2 of \cite{NeumannWangerin1894}), while all three-phase contacts involving the solid substrate are governed by a Young law. When studying the relaxational dynamics of such a system any dynamic model needs to be able to dynamically recover the possible equilibrium states as determined by the macroscopic parameters.

The full dynamic behavior of a two-layer liquid film, a drop on a liquid substrate or a compound drop can be studied with a thin-film (aka long-wave or lubrication) model that couples dynamic equations for the two interface profiles \cite{OrDB1997rmp,PBMT2005jcp,CrMa2009rmp} as done in many of the above-given references. {Such mesoscopic hydrodynamic models} incorporate wettability via a wetting energy (related to the disjoining pressures in the films) and, in the case of partial wetting, allow for the coexistence of large drops of finite contact angle with thin adsorption layers, i.e., sharp three-phase contact lines are replaced by smooth transition regions.

Ref.~\cite{DiTh2025prf} then notes that to cover the full parameter space spanned by the six macroscopic interface energies, the earlier existing formulations for the wetting energies have to be amended. In consequence, they derive {conditions that ensure consistency between macroscopic and mesoscopic descriptions that allow one to directly relate macroscopic and mesoscopic parameters (selected interface energies and parameters controlling the wetting energies, respectively). Their resulting mesoscopic hydrodynamic model} can be employed in a long-wave and a full-curvature version (the merit of and argument behind the latter is discussed in \cite{DiTh2025prf} and, more generally, in \cite{Thie2018csa}).

{After developing the model, \cite{DiTh2025prf} mainly studies spreading compound drops and coarsening ensembles of such drops on horizontal substrates. However, sliding compound drops are only briefly considered. Here, we employ the mesoscopic hydrodynamic model in the full-curvature variant presented in \cite{DiTh2025prf}, and focus on sliding compound drops that consist of two immiscible non-volatile partially wetting liquids on an inclined homogeneous smooth solid substrate. The particular aim of the performed parameter study is the identification of qualitative transitions where at specific values of the considered control parameters (driving force, ratios of drop volumes and viscosities) discontinuous transitions between different drop configurations and time-periodic behavior occur. Besides on the underlying bifurcation structure our interest focuses on the quantitative changes in the properties of the sliding drops in the different configurations. Our analysis is based on a combination of numerical time simulations \cite{HeHa2006} and path continuation \cite{UeWR2014nmma,EGUW2019springer, DiRo2024jocp}, and furthermore includes a discussion of the different dissipation channels.}

{The work is structured as follows. Section~\ref{sec:model} briefly introduces the employed full-curvature variant of the mesoscopic hydrodynamic model, for a full account we refer to Ref.~\cite{DiTh2025prf}. Subsequently, section~\ref{sec:drop-on-layer} investigate the reference case of a drop of liquid on a layer of another liquid that acts as an adaptive substrate. Section~\ref{sec:drifting_states} then analyzes properties of and transitions between stationary sliding compound drops in different configurations. %
This is followed in section~\ref{sec:transients} by a brief discussion of the time-periodic states corresponding to cyclic fusion-overtaking-splitting dynamics that occur beyond the parameter range where stable stationary sliding compound drops exist.} Finally, we conclude and give an outlook in section~\ref{sec:conc}.

\section{The mesoscopic model}
\label{sec:model}
\subsection{Governing equations}
\label{subsec:gov-eq}
\begin{figure*}[t]
	\centering
	\includegraphics[width=0.75\linewidth]{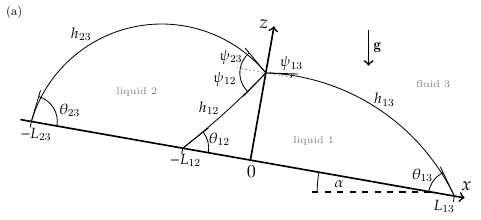}
	\includegraphics[width=0.75\linewidth]{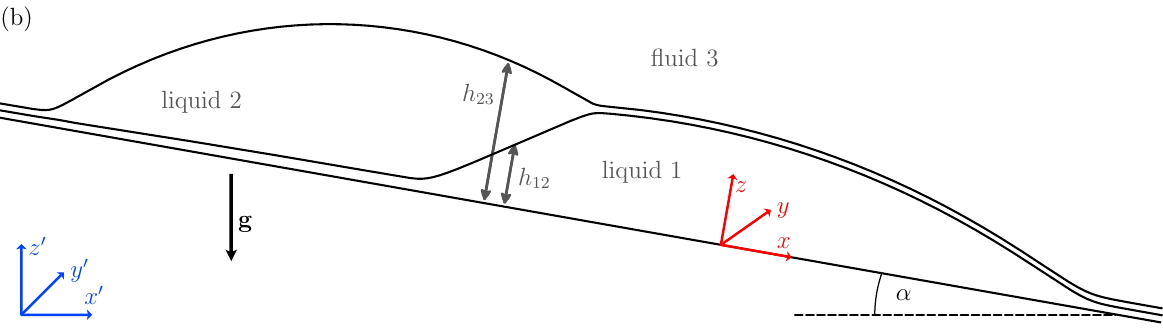}
	\caption{Sketches of the considered geometry showing as an example an asymmetric compound drop in 2-1 configuration (drop of liquid~1 in front of drop of liquid~2) on an inclined plane in the (a) macroscopic and (b) mesoscopic picture. The substrate is tilted by an angle $\alpha$ w.r.t.\ the horizontal $x^\prime$-direction, yielding a rotation between the laboratory system (dashed coordinates, blue) and the substrate coordinates (coordinates without dashes, red). Gravity is acting in the negative $z^\prime$-direction. The macroscopic description allows for direct Young and Neumann constructions at the three-phase contacts. The mesoscopic description features various adsorption layers and contact angles are determined at the inflection point of the respective profile (see Fig.~1 in~\cite{DiTh2025prf}).
       }
	\label{fig:th_model_sketch}
      \end{figure*}

To model sliding compound drops we employ the mesoscopic hydrodynamic model of Ref.~\cite{DiTh2025prf} -- a two-layer thin-film model that amends the description in \cite{PBMT2005jcp} by incorporating wetting energies that allow for a complete mapping of the relevant parameter space of macroscopic interface energies, for a discussion see sections~II and III.A of \cite{DiTh2025prf}.
The dynamic equations are written as a gradient dynamics for two conserved order parameter fields \cite{Thie2018csa}, namely, the height profiles of the interface between liquids~1 and~2 $\hOT(\vec{r},t)$ and of the interface between liquid~2 and the ambient gas (medium~3) $\hTT(\vec{r},t)$ (see sketches in Figs.~\ref{fig:th_model_sketch}(a) and \ref{fig:th_model_sketch}(b) for the macroscopic and mesoscopic picture, respectively). {The kinetic equations are}
\begin{equation}
	\label{eq:thinFilm-interfaceHeights}
{\small
	\begin{aligned}	
		\partial_t \hOT &= -\nabla \cdot \vec{j}_{12} = \nabla \cdot  \left(Q_{11}\nabla\dFdhOT + Q_{12}\nabla\dFdhTT\right),\\
		\partial_t \hTT &= -\nabla \cdot \vec{j}_{23} = \nabla \cdot  \left(Q_{21}\nabla\dFdhOT + Q_{22}\nabla\dFdhTT\right),\\
	\end{aligned}
}
\end{equation}
where $\mathcal{F}$ denotes the underlying energy functional, and the symmetric and positive definite mobility matrix is
\begin{equation}
	\label{eq:Mobility_interfaceHeights}
{\tiny 	
	\tens{Q} = 
  \frac{1}{\eta_1}
	\begin{pmatrix}
		\frac{\hOT^3}{3} & \frac{\hOT}{2}\left(\hTT-\frac{\hOT}{3}\right)\\
		\frac{\hOT}{2}\left(\hTT-\frac{\hOT}{3}\right) & \frac{(\hTT-\hOT)^3}{3}\left(\frac{\eta_1}{\eta_2} - 1\right) + \frac{\hTT^3}{3}
	\end{pmatrix}
}
\end{equation}
as derived using long-wave approximation from the Navier-Stokes equations with no-slip boundary conditions at the solid substrate, stress and velocity continuity at the liquid~1-liquid~2 interface, stress-free liquid~2-gas interface, and kinematic boundary conditions at liquid~1-liquid~2 and liquid~2-gas interfaces \cite{PBMT2005jcp}.
Furthermore, $\eta_1$ and $\eta_2$ are the dynamic viscosities of the respective liquids, $\partial_t$ is the partial derivative w.r.t.\ time $t$, the nabla operator is $\nabla = (\partial_x, \partial_y)^T$, where $\vec{r}=(x,y)^T$ are the coordinates along the two-dimensional planar rigid solid homogeneous substrate.

For a horizontal substrate the underlying mesoscopic energy functional reads
\begin{equation}
	\label{eq:F_horizontal}
\begin{split}
	\mathcal{F} = \mathcal{F}_{\mathrm{h}} = \int_A [\gamSOne + \gamOneTwo\xi_{12} + \gamTwoThr\xi_{23} \\
  + g(h_{12},h_{23},\xi_{23})] \, \mathrm{d}^2r,
\end{split}
\end{equation}
where $\gamSOne$, $\gamOneTwo$, and $\gamTwoThr$ are the interface energy densities for the solid-liquid~1, liquid~1-liquid~2, and liquid~2-gas interface, respectively, while the final term represents the overall wetting energy. In particular, we use the combination
\begin{equation}
	\label{eq:wetting}
\begin{split}
	g(h_{12},h_{23},\xi_{23}) &= f_1(\hOT) + \xi_{23}f_2(\hTT - \hOT) \\
	& ~~~~ + f_3(\hTT)
\end{split}
\end{equation}
with terms corresponding to the wettability of liquid~1 on the substrate under liquid~2 as surrounding medium, of liquid~2 on liquid~1, and liquid~2 on the substrate under the gas phase. The wettability of liquid~1 on the substrate under the gas phase results from the combination of the first and third terms. Each of the three terms includes long-range attracting and short-range repelling forces, {i.e., $f_i(h) = A_i/(2h^{2})\left[2/5(h_{i}^\star / h)^{3} - 1\right]$, where $h$ is the respective argument, $A_i$ the corresponding Hamaker constant \cite{Israelachvili2011}, and $h_{i}^\star$ are parameters related to the adsorption layer thicknesses. In the case of partial wetting, the minima $f(h_i^\star) = -3A_i/10(h_i^\star)^{2}$ at $h=h_i^\star$ define equilibrium adsorption layer thicknesses, i.e., $h_1^\star = h_{\mathrm{a}1}$, $h_2^\star = h_{\mathrm{a}2}$ and $h_3^\star = h_{\mathrm{a}1} + h_{\mathrm{a}2}$. Note that combinations of $\gamSOne$, $\gamOneTwo$, $\gamTwoThr$, and of the minimal values of the wetting energies give all six occurring macroscopic interface energies via consistency relations that quantitatively relate the macroscopic and the mesoscopic descriptions. Explicitly, they read $\gamSTwo = \gamSOne + \gamOneTwo + f_1(h_{\mathrm{a1}})$, $\gamSThr = \gamSOne + \gamOneTwo + \gamTwoThr + f_1(h_{\mathrm{a1}}) + f_2(h_{\mathrm{a2}}) + f_3(h_{\mathrm{a1}} + h_{\mathrm{a2}})$, and $\gamOneThr = \gamOneTwo + \gamTwoThr + f_2(h_{\mathrm{a2}})$. Inserting back the minimal values into these equations directly relates the macroscopic interface energies to the three Hamaker constants $A_i$. For further} details see sections~III.A to III.C of \cite{DiTh2025prf}.

In Eqs.~\eqref{eq:F_horizontal} and~\eqref{eq:wetting} the functions $\xi_{ij}$ represent metric factors, either in their long-wave form or, as used here, in their exact form $\xi_{12}=\sqrt{1 + |\nabla \hOT|^2}$ and $\xi_{23}=\sqrt{1 + |\nabla \hTT|^2}$. The latter results in the full-curvature version of mesoscopic hydrodynamics, for a discussion see \cite{Thie2018csa,DiTh2025prf}. In particular,  \cite{DiTh2025prf} shows that the metric factor in Eq.~\eqref{eq:wetting} is needed to faithfully recover Neumann's law.

Next, we incorporate the potential energy of gravity $\FG$ in addition to the discussed $\mathcal{F}_{\mathrm{h}}$, and model sliding compound drops on an inclined substrate. For a two-dimensional substrate inclined in $x$-direction we have
\begin{equation}
	\label{eq:FG_LabFrame}
	\FG = \int_{V_1} \rho_1 g z^\prime \, \mathrm{d} z^\prime \mathrm{d}^2r^\prime + \int_{V_2} \rho_2 g z^\prime \, \mathrm{d} z^\prime\mathrm{d}^2r^{\prime},
\end{equation}
where $g$ is the gravitational acceleration and $\rho_i$ and $V_i$ are the mass density and the volume of liquid $i$, respectively. In the laboratory frame (with dashes, shown in blue in Fig.~\ref{fig:th_model_sketch}), gravity acts in the negative $z^{\prime}-$direction. To include the effect into Eq.~\eqref{eq:thinFilm-interfaceHeights}, a coordinate transformation is carried out into the frame of the tilted substrate (without dashes, shown in red in Fig.~\ref{fig:th_model_sketch}). The corresponding rotation matrix is
\begin{equation}
	\tens{R}(\alpha) = 
	\begin{pmatrix}
		\cos(\alpha) & 0& -\sin(\alpha)\\
		0 & 1 & 0\\
		\sin(\alpha) & 0& \cos(\alpha)
	\end{pmatrix},
\end{equation}
where $\alpha$ is the inclination angle.
As a result Eq.~\eqref{eq:FG_LabFrame} becomes
\begin{align}
	\begin{split}
	\FG &= \int_A \int_0^{\hOT} \rho_1 g (z\cos\alpha + x\sin\alpha) \, \mathrm{d}z\,\mathrm{d}^2r \\
	&~~~~~ + \int_A \int_{\hOT}^{\hTT} \rho_2 g (z\cos\alpha + x\sin\alpha) \, \mathrm{d}z\,\mathrm{d}^2r
	\end{split}\\
	\begin{split}
	&= g\int_A \left[\frac{1}{2}(\rho_1-\rho_2)\hOT^2\cos\alpha + \frac{1}{2}\rho_2\hTT^2\cos\alpha \right. \\ 
	&~~~~~ \left.+ (\rho_1-\rho_2)\hOT x\sin\alpha  + \rho_2\hTT x\sin\alpha\right] \,  \mathrm{d}^2r.
	\end{split}
\end{align}
Here, we are mainly interested in the sliding behavior of drops due to the laterally acting forces  encoded in the terms $\sim \sin\alpha$, but not in the flattening influence of the hydrostatic forces (terms $\sim \cos\alpha$) that we neglect in the following. In other words, we only consider either small slowly sliding drops or drops on vertical substrates with $\alpha = \pi/2$. To capture both cases we introduce an effective inclination parameter $\beta=\rho_2 g \sin\alpha$ and the relative density $\rho_\mathrm{r}=\rho_1/\rho_2-1$ and obtain for the overall energy functional
\begin{equation}
  \mathcal{F} = \mathcal{F}_\text{h} + \int_A (\rho_\mathrm{r} \hOT x + \hTT x) \beta \, \mathrm{d}^2r,
  \label{eq:ff2}
\end{equation}
with the pressure jumps at the two interfaces obtained as functional derivatives of $\mathcal{F}$ as
\begin{align}
	\label{eq:dFdh12}
	\begin{split}	
	P_1 - P_2 &= \dFdhOT\\ 
	&= -\gamOneTwo\frac{\nabla^2\hOT}{\xi_{12}^3} + f_1^\prime - \xi_{23}f_2^\prime + \rho_\mathrm{r} \beta x,
	\end{split}\\ 
	\label{eq:dFdh23}
	\begin{split}	
	P_2 - P_\mathrm{g} &= \dFdhTT \\
	&= -(\gamTwoThr+f_2)\frac{\nabla^2\hTT}{\xi_{23}^3} + f_3^\prime\\
	&~~~~ + \frac{1}{\xi_{23}}\left(1+(\nabla \hOT)\cdot(\nabla\hTT)\right)f_2^\prime + \beta x.
	\end{split}
\end{align}
where in the following we set the pressure in the ambient gas $P_\mathrm{g} $ to zero.
Here, $f_i^\prime$ stands for the derivative w.r.t.\ the respective argument. Note that the expression $\nabla^2h_{ij}/\xi_{ij}^3$ corresponds to the exact curvature.

\subsection{Nondimensionalization, parameters, and numerical methods}
\label{subsec:nondim-parameters-setup}
We effectively nondimensionalize the dynamic equations by setting $\eta_1=\gamma_{12}=h_\mathrm{a1}=1$, see {appendix \ref{sec:app_nondim}}. Furthermore, we use $\gamma_\mathrm{s1}=0$ as only differences of substrate energies are relevant. The remaining free parameters are the macroscale interface energies that we fix as $(\gamOneThr, \gamTwoThr, \gamSTwo, \gamSThr) = (1.6,0.8,0.85,1.4)$ resulting in the Hamaker constants ${(A_1,A_2,A_3) \approx (0.5,0.67,0. 67)}$, and the macroscopic equilibrium contact angles $(\vartheta_{12},\vartheta_{13},\vartheta_{23}) \approx (\SI{31,8}{\degree},\SI{29,0}{\degree}, \SI{46,6}{\degree})$. 
Using identical mass densities, i.e., $\rho_1 = \rho_2 =\rho$ gives $\rho_\mathrm{r} = 0$ (also cf.~\cite{DiTh2025prf,HDGT2024l}) simplifying the free energy to $\mathcal{F} = \mathcal{F}_\text{h} + \int_A \beta\hTT x \, \mathrm{d}^2r$.

The remaining free parameters represent our control parameters in the following sections, i.e., the inclination parameter $\beta$, the viscosity ratio $\VisVer = \eta_2/\eta_1$ and either the mean thickness of the layer of liquid~1 that acts as an adaptive substrate for a sliding drop of liquid~2 (section~\ref{sec:drop-on-layer}) or the volume ratio $\VolVer = V_2/V_1$  for sliding compound drops (section~\ref{sec:drifting_states}). We consider a one-dimensional domain of fixed size $L=1000$, if not stated otherwise, set up symmetrically around $x=0$ with periodic boundary conditions. As a solution measure we use the L$_2$-norm of the liquid~2-gas~interface profile defined as $\left\lVert\hTT\right\rVert_2 = \sqrt{\int_{-L/2}^{L/2} \lvert\hTT\rvert^{2}\, \mathrm{d}x}$.

The governing equations are solved numerically employing time integration and pseudo-arclength continuation. For time simulations we use the finite-element library \textit{oomph-lib} with its space and time adaptivity \cite{HeHa2006} while the bifurcation diagrams are obtained using the continuation toolbox \textit{pde2path}~\cite{UeWR2014nmma} {and confirmed by \textit{pyoomph}~\cite{DiRo2024jocp} that builds on \textit{oomph-lib} and \textit{GiNaC}~\cite{BaFK2002josc}. Furthermore, two-parameter continuations are performed with \textit{pyoomph}.}

\section{Drop sliding on liquid substrate}
\label{sec:drop-on-layer}

\begin{figure*}
	\centering
	\includegraphics[width=\linewidth]{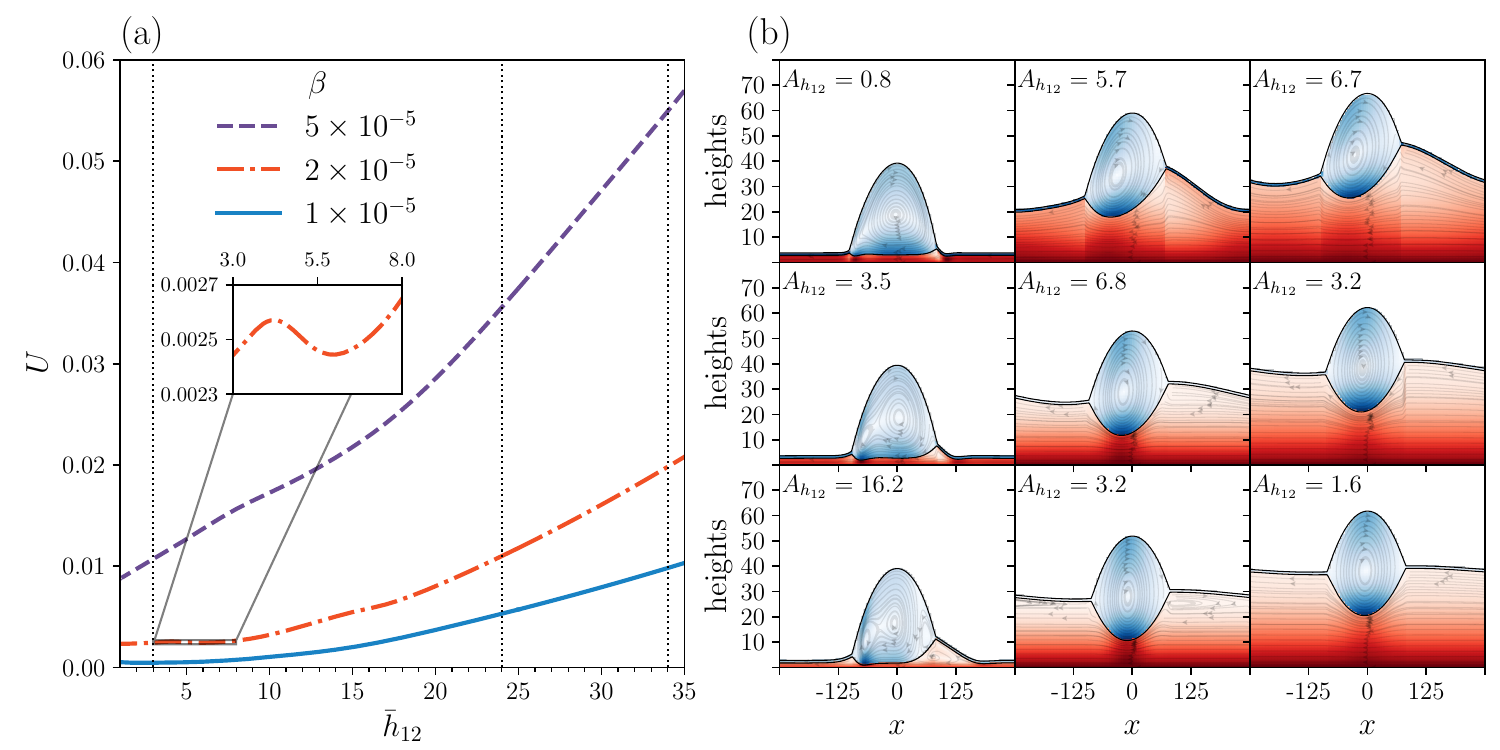}
	\caption{Behavior of a drop of liquid~2 sliding stationary down on a layer of liquid~1. Panel (a) shows the dependence of the drop velocity $\vel$ on the mean thickness of the liquid substrate (liquid~1) $\bar h_{12}$ for a number of different inclination parameters $\beta$ as given in the legend. Panel (b) gives selected corresponding drop profiles at the $\bar{h}_{12}$-values marked in (a) by thin vertical lines (with the same left-to-right order), including streamlines and the color-encoded absolute value of the velocity field (both in the comoving frame).  The vertical order of the profiles in (b) mirrors the vertical order of the curves in (a), i.e., the top [bottom] row are for the largest [smallest] $\beta$. In each panel the value of an asymmetry measure is given. It is defined as $A_{h_{12}}=\frac{1}{2V_1}\int xh_{12}\mathrm{d}x$ (cf.~Ref.~\cite{HeST2021sm}), i.e., here, based on the profile of the liquid substrate. The domain size is $L=500$, the mean thickness of the upper layer is $\bar{h}_{23}-\bar{h}_{12}=10$, and its viscosity is $\eta_2=1$. All other parameters are as described in section~\ref{subsec:nondim-parameters-setup}.
	}
	\label{fig:drop-on-layer}
\end{figure*}

First, we briefly investigate the dynamic behavior of a drop of liquid~2 that partially wets a layer of liquid~1 that itself ideally wets the solid substrate, i.e., it acts as an adaptive substrate for the sliding drop of liquid~2. The situation is similar to the one studied in Refs.~\cite{CrMa2006jcis,JHKP2013sjam,HJKP2015jem} for horizontal substrates, i.e., with a focus on spreading and equilibrium shapes of drops on a liquid substrate, and in Ref.~\cite{KrMi2003sjam} for drops on thick inclined liquid layers.

Here, we use the model introduced in section~\ref{subsec:gov-eq} omitting two of the three terms in the wetting energy, i.e., we use $f_1=f_3=0$ in Eq.~\eqref{eq:wetting}, and only keep the part responsible for the wettability of liquid~2 on liquid~1 (that is, $f_2$). Fig.~\ref{fig:drop-on-layer}(a) presents the dependence of the drop velocity $\vel$ on the mean thickness of the liquid substrate $\bar h_{12}$ for different inclination parameters $\beta$. 

Fig.~\ref{fig:drop-on-layer}(b) gives selected corresponding drop profiles together with the global asymmetry measure $A_{h_{12}}$ (defined in the caption), with corresponding velocity fields (shown as streamlines and absolute value of velocity field). The curves and profiles are obtained by path continuation and are furthermore confirmed by numerical time simulations (not shown). In the former method the drop velocity is determined as nonlinear eigenvalue while in the latter it is extracted from the obtained data by averaging the center-of-mass (CoM) velocity of the upper liquids over a small time span. 

Inspecting Fig.~\ref{fig:drop-on-layer}(a) we observe that at first sight $\vel$  seems to increase monotonically but nonlinearly with $\bar h_{12}$ for all three shown $\beta$-values. At {moderate  $\bar h_{12}$ an asymmetric liquid lens slides down on a thick film with advancing angles that are larger than the receding ones (cf.~Fig.~\ref{fig:drop-on-layer}(b) central column for $\bar{h}_{12}=24$), and with a velocity that scales approximately linear with $\bar h_{12}$.  Note, however, that the asymmetry measure $A_{h_{12}}$ changes nonmonotonically with increasing $\beta$: At small to moderate $\beta$,  the value $A_{h_{12}}$ increases as the sliding drop creates a slope of the liquid~1-gas interface on a large scale. However, at larger $\beta$ the slope of $h_{12}$ becomes increasingly localized in front of the drop resulting in a decrease of  $A_{h_{12}}$. 

Interestingly, {the behavior qualitatively differs at small and large driving force. At small $\beta$ the asymmetry $A_{h_{12}}$ of the profile increases with decreasing $\bar h_{12}$ while it decreases at large $\beta$. In consequence, for very thin liquid substrates (i.e., fixed small  $\bar h_{12}$) the asymmetry decreases with increasing $\beta$ in the shown $\beta$-range} (cf.~Fig.~\ref{fig:drop-on-layer}(b) left column for $\bar{h}_{12}=3$).  Note, for instance, the relatively pronounced wetting ridges for $\beta=1\times10^{-5}$ and $\beta=2\times10^{-5}$ that are larger at the advancing edge of the drop than at the receding edge. However, as for $\beta=0$ the drops are perfectly symmetric ($A_{h_{12}}=0$) our observation implies that for the lower $\bar{h}_{12}$-values the maximal asymmetry occurs for $\beta<1\times10^{-5}$. Furthermore, a close inspection of the $\vel(\bar{h}_{12})$-dependence for $\beta=2\times10^{-5}$ reveals that the velocity changes nonmonotonically (cf.~inset of Fig.~\ref{fig:drop-on-layer}(a)). 

{In contrast, for} very thick liquid substrates  (i.e., $\bar{h}_{12}=34$, right column in panel~(b)), we observe {the reverse trend for} the asymmetry measure.  There, $A_{h_{12}}$ increases with increasing inclination, and we expect the $A_{h_{12}}$ to peak at larger $\beta$. Thus, the stationary state of maximal asymmetry is found for large [moderate, small] liquid substrate thicknesses~$\bar{h}_{12}$ at large [moderate, small]~$\beta$.

{The behavior can be explained by a qualitative argument: When the friction forces are small, i.e., at small driving force and small drop velocity, the liquid substrate is able to adapt more strongly to the capillary forces exerted by the sliding drop. This results in the larger wetting ridges at smaller $\beta$. In contrast, at larger $\beta$ the friction forces due to shear stress are larger than the capillary forces in the contact line regions resulting, e.g., in a much smaller wetting ridge in the top left panel of  Fig.~\ref{fig:drop-on-layer}(b) than in the bottom left panel. However, due to the parabolic velocity profile in the liquid substrate, the effect of shear stress gets smaller with increasing substrate thickness. In consequence, for thick substrates (right column of Fig.~\ref{fig:drop-on-layer}(b)) the effect of the dynamic contact angles dominates.}

{After considering the behavior} of  the sliding drop on a liquid substrate {that we will come back to later on}, we next turn to our main focus, {a parameter study of sliding compound drops.} In contrast to the present section we return to the full form of the wetting energy given in Eq.~\eqref{eq:wetting}. Hence, liquids~1 and~2 both partially wet the substrate, and liquid~2 also partially wets liquid~1.
           
\section{Stationary sliding compound drops}
\label{sec:drifting_states}

Next, we investigate the properties of stationary sliding compound drops that always exist for sufficiently small substrate inclination $\beta$ as every steady asymmetric drop that exists for horizontal substrate \cite{DiTh2025prf} spawns a branch of sliding drops {parametrized by substrate inclination or driving force.} At larger $\beta$ the branch might undergo various bifurcations, {e.g., it may ceases to exist at a saddle-node bifurcation or become more and more unstable in a sequence of such bifurcations.}

Here, we consider the dependence of the {properties of the compound drop} on a number of relevant control parameters, namely, the inclination parameter $\beta$ (section~\ref{subsec:var_of_inclination}) {and the volume ratio $\VolVer$ (section~\ref{subsec:var_of_volume})} at fixed values of the remaining parameters. {Additionally, the influence of the viscosity ratio is analyzed in appendix~\ref{sec:app_var_of_viscosity}.} As in section~\ref{sec:drop-on-layer}, {numerical time simulations and path continuation are employed to obtain the drop velocity $\vel$ (here, as the mean of the CoM velocities of the two individual liquids).} Furthermore, we determine the dynamic angles at all three-phase contacts, i.e.,  all dynamic Young and Neumann angles. In particular, we obtain the angles from the slopes at the inflection points of the {profiles obtained by time simulation}. Note that we use Neumann angles defined as the angles to the closest horizontal line (see Fig.~\ref{fig:th_model_sketch}) as {this makes it possible to present all of them in a single figure that focuses on a relatively small range of angles.}
\subsection{Influence of the inclination parameter}
\label{subsec:var_of_inclination}
\begin{figure*}
	\centering
	\includegraphics[width=0.75\linewidth]{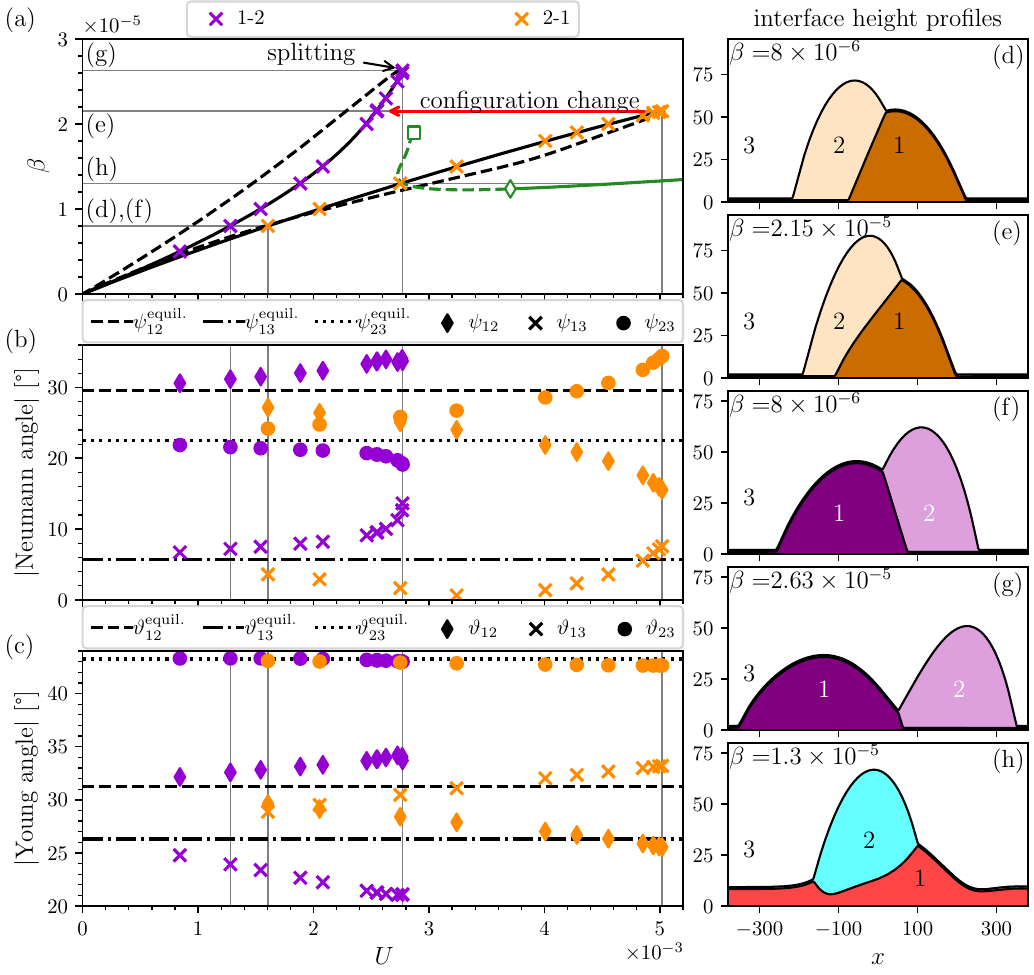}
	\caption{Characterization of stationary sliding compound drops in dependence of the inclination parameter $\beta$. Panel (a) shows the drop velocity $\vel$ (horizontal axis) as a function of $\beta$ (vertical axis) while (b) and (c) show the absolute values of the three Neumann angles and the three Young angles, respectively (see legends), all as a function of $\vel$. In (a), solid [dashed] black lines indicate stable [unstable] states obtained by continuation.  {Compound drops in 2-1 and 1-2 configuration obtained by time simulation are marked by orange and purple cross symbols, respectively. Note that also drops sliding on a liquid layer exist shown in panel (a) as a green line. The unstable part of the branch is only followed till the green square symbol. Panels (d,e), (g,f), and (h) give example profiles for 2-1, 1-2, and drop-on-layer configurations, respectively. The corresponding values of $\beta$ and $\vel$ are indicated in panels (a-c)} by thin solid vertical and horizontal lines. The dotted, dashed and dot-dashed lines in (b) and (c) mark the various equilibrium angles (see legends). The volume ratio and the viscosity ratio are fixed to one, $\VolVer=\VisVer=1$. The remaining parameters are given in section~\ref{subsec:nondim-parameters-setup}.}
	\label{fig:plot_beta_variation}
\end{figure*}

First, the inclination parameter $\beta$ is varied at fixed $\VolVer=\VisVer=1$, see Fig.~\ref{fig:plot_beta_variation}. The dependence of compound drop velocity $\vel$ on the control parameter $\beta$ is shown in Fig.~\ref{fig:plot_beta_variation}(a). There, solid and dashed lines indicate branches of stable  and unstable sliding drops, respectively, as obtained by path continuation. In contrast, cross symbols represent data obtained from time simulations. The corresponding absolute values of Neumann and Young angles are shown as a function of $\vel$ in Figs.~\ref{fig:plot_beta_variation}(b) and \ref{fig:plot_beta_variation}(c), respectively.\footnote{Note that due to the display of the absolute values, changes in sign of a corresponding angle look like a \enquote{kink} at zero angle (see e.g.\ orange crosses in Fig.~\ref{fig:plot_beta_variation}(b) at $\vel\approx 3.5\times10^{-3}$).} A selection of profiles of stable stationary sliding drops is presented in Figs.~\ref{fig:plot_beta_variation}(d)-(g). The selected $\beta$-values are marked by thin horizontal lines in Fig.~\ref{fig:plot_beta_variation}(a).

At the chosen set of energetic parameters, the asymmetric compound drop is for a horizontal substrate ($\beta=0$) the stable state of lowest energy. Due to isotropy of space it exists in two equivalent variants, i.e., 1-2 and 2-1 configuration (these labels corresponds to the lateral order of the liquid drops, from left to right), in the following indicated by orange and purple coloring, respectively. They are related by reflection. However, for $\beta>0$ the spatial symmetry is broken and the two variants give rise to the two branches of linearly stable sliding compound drops {shown in Fig.~\ref{fig:plot_beta_variation}(a) as solid black lines.} Besides these states, that are most relevant here, at $\beta=0$ also a stable symmetric drop-on-drop state exists (cf.~\cite{DiTh2025prf}) that is found as an equilibrium state in time simulations. However, it is rendered asymmetric and unstable for any finite $\beta$. Here, we do not further discuss it.

We note that over the entire range where the {asymmetric stationary sliding drops} exist, their velocity monotonically increases with increasing inclination parameter $\beta$. This is in line with expectations, as an increasing $\beta$ implies a larger potential energy that has to be dissipated per passed substrate length. In other words, the monotonic behavior indicates that {for a given drop configuration} the dissipation also increases monotonically with drop velocity. For an analysis of the spatial distribution of dissipation see below.

At small $\beta$ the slopes of the 1-2 and 2-1 branches in Fig.~\ref{fig:plot_beta_variation}(a) are identical {as can be concluded from the symmetry (1-2 state, $\beta$,~$\vel$) $\to$ (2-1 state, $-\beta$,~$-\vel$).} However, with increasing $\beta$ the branches separate from each other - at identical $\beta$ the drop in 2-1 configuration moves by up to about a factor two faster than the one in 1-2 configuration. For drops in the 2-1 configuration the Young angles behave as expected: The advancing angle ($\vartheta_{13}$) increases with increasing velocity, while both receding ones ($\vartheta_{12}$ and $\vartheta_{23}$) decrease (Fig.~\ref{fig:plot_beta_variation}(c)). The change in Neumann angles at {small and moderate} $\beta$ results from a solid body-like rotation of the entire Neumann region.\footnote{While all absolute Neumann angles change by several degrees, the residues of the Neumann law remain on a small finite constant level. Macroscopically these residues would be zero, but in the mesoscopic description they remain finite due to the way the angles are measured. Further details are discussed in Ref.~\cite{BT_Thy_2024}.} At larger $\beta$ the angles change individually and the residuals strongly increase. Overall, the behavior of the drop in the 1-2 configuration is similar: Now, $\vartheta_{23}$ and $\vartheta_{12}$ are advancing angles but only the latter clearly increases while $\vartheta_{23}$ slightly decreases. The receding angle $\vartheta_{13}$ decreases with increasing $\vel$.

Overall, the most notable feature of the bifurcation diagram in Fig.~\ref{fig:plot_beta_variation}(a) is that both configurations cease to exist at respective critical values of $\beta$ (and in consequence of $\vel$). In particular, when increasing $\beta$ from zero, first the drop in 2-1 configuration (that moves faster than the one in 1-2 configuration) ceases to exist at a {saddle-node bifurcation occurring at a} critical $\beta_{\mathrm{c}21} \approx 2.15\times10^{-5}$ (corresponding to a critical velocity $\vel_{\mathrm{c}21} \approx 5.02\times10^{-3}$). {There, the branch folds back towards smaller $\beta$ implying that the states on the dashed part are unstable.}

{When starting with a 2-1 drop and increasing $\beta$ above $\beta_{\mathrm{c}21}$, the trailing drop of liquid 2 slides over the drop of liquid 1 and a stationary 1-2 drop results (also see movie~1 in the Supplemental Material). The dynamic configuration change is indicated in Fig.~\ref{fig:plot_beta_variation}(a) by a red arrow.} 
Interestingly, during the overtaking process the compound drop nearly halves its velocity, i.e., decreases its dissipation per time. Fig.~\ref{fig:plot_beta_variation}(b) shows that the Neumann angles strongly change when $\beta_{\mathrm{c}21}$ is approached while the Young angles are less affected and show a linear trend (see Fig.~\ref{fig:plot_beta_variation}(c)). The height profiles in Figs.~\ref{fig:plot_beta_variation}(d) and \ref{fig:plot_beta_variation}(e) suggest the Neumann region strongly rotates.

Further increasing $\beta$ beyond $\beta_{\mathrm{c}21}$, the {1-2 drop} continues to increase its velocity until also this state ceases to exist at a {saddle-node bifurcation that occurs at a} second critical inclination $\beta_{\mathrm{c}12} \approx 2.63\times10^{-5}$ (corresponding to $\vel_{\mathrm{c}12} \approx 2.77\times10^{-3}$). At the bifurcation} the Neumann region approaches the substrate, see profiles in Figs.~\ref{fig:plot_beta_variation}(f) and \ref{fig:plot_beta_variation}(g). %
{This is consistent with the observation that beyond the bifurcation a compound drop splits into two distinct drops that move with individual velocities. In a setting with periodic boundaries or with a periodic array of drops, a regular time-periodic process emerges that is further discussed in section~\ref{sec:transients} (see movie~2 in the Supplemental Material).}
{The Neumann angles also show a strongly nonlinear behavior when approached this second dynamic transition} (Fig.~\ref{fig:plot_beta_variation}(b)) while the change  in the Young angles remains approximately linear (Fig.~\ref{fig:plot_beta_variation}(c)).

{Depending on initial conditions, at large driving force one finds two stable states: the already mentioned time-periodic splitting-fusion cycle and, alternatively, a drop stationary sliding on a liquid layer, similar to the states in section~\ref{sec:drop-on-layer}. The latter states form a branch in the bifurcation diagram represented by  the green line in Fig.~\ref{fig:plot_beta_variation}(a).\footnote{The stable part of the branch  extends monotonically toward larger inclinations (and velocities). The unstable part of the branch is only followed till the green square symbol.}  An exemplary profile is shown in Fig.~\ref{fig:plot_beta_variation}(h). Note that in contrast to  section~\ref{sec:drop-on-layer}, here, these states are computed including all terms of the wetting energy \eqref{eq:wetting}, i.e., the sliding drop dynamically stabilizes the dewetting instability of the liquid substrate layer. In situations where an initially uniform two-layer film undergoes dewetting on the inclined substrate, this states naturally emerge if the bottom layer dewets sufficiently slower than the upper layer (see movie~3 in the Supplemental Material). Interestingly, when following the green branch in Fig.~\ref{fig:plot_beta_variation}(a) with decreasing velocity (and inclination), it is first destabilized at a Hopf bifurcation at $(\beta_{\mathrm{H}}, \vel_{\mathrm{H}}) \approx (1.236\times 10^{-5}, 3.7\times 10^{-3})$ (marked by a diamond symbol)\footnote{For completeness we mention that the Hopf bifurcation is supercritical and in its close vicinity (slightly decreasing $\beta$) time simulations show a stable oscillatory state where the minimal thickness of the bottom layer oscillates in time, i.e., the interaction of the sliding drop with the dewetting mode results in a stable surface wave of the liquid substrate. If one moves further beyond the bifurcation the dewetting mode dominates and the system again converges to a sliding compound drop.} before turning back toward larger $\beta$ at a saddle-node bifurcation at $(\beta_{\mathrm{c}}, \vel_{\mathrm{c}}) \approx (1.225\times 10^{-5}, 3.3\times 10^{-3})$.}

\begin{figure*}
	\centering
	\includegraphics[width=\linewidth]{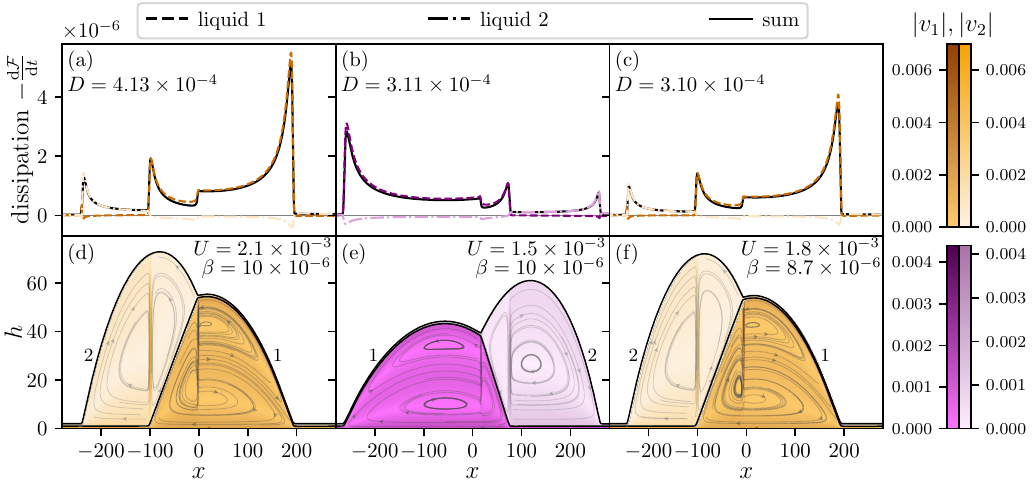}
	\caption{(a)-(c) Vertically integrated dissipation and (d)-(e) velocity fields of the stationary sliding compound drops in (left and right) 2-1 configuration and (middle) 1-2 configuration. Panels (a)-(c) give the laterally spatially resolved dissipation for the two liquids individually and the total value along with the integrated dissipation $D$. Panels (d)-(e) give the height profiles together with the streamlines and the color-coded magnitude of the velocity fields (both in the comoving frame). Note that the velocity $\vel$ of the compound drop (and thus the comoving frame) differs for each column. Datasets in the two leftmost and rightmost columns are obtained at the same inclination parameter $\beta=10^{-5}$ and integrated dissipation $D\approx3.1\times10^{-4}$, respectively (see labels). Remaining parameters are as in section~\ref{subsec:nondim-parameters-setup}.}
	\label{fig:plot_Dissipation_cd}
\end{figure*}

{To obtain a better physical understanding of the different velocities of the compound drops in different configurations at identical $\beta$, we next analyze the corresponding dissipation profiles. Fig.~\ref{fig:plot_beta_variation}(a) indicates that as the given parameters a 2-1 drop always moves faster than a 1-2 drop. Appendix~\ref{sec:app_single_droplets}  shows that at identical inclination, volume and viscosity an individual drop of liquid~1  moves (with e.g.\ $\vel\approx1.3\times10^{-3}$ at $\beta=10^{-5}$) by about a factor~6 slower than one of liquid~2. The velocities of both compound drops at this $\beta$ (Fig.~\ref{fig:plot_beta_variation}) are relatively close to the lower value. This indicates that the slow drop dominates the behavior of the compound drop for both configurations, i.e., the system behavior is determined by its slowest element.}

To discuss this {more quantitatively, Fig.~\ref{fig:plot_Dissipation_cd} compares the dissipation of the two configurations} on the one hand at equal inclination parameter $\beta=10^{-5}$ (left and center column) and on the other hand at equal total dissipation $D\approx3.1\times10^{-4}$ (center and right column). For details {on} the calculation of the dissipation profiles {within the two liquids,} see appendix~\ref{sec:app_diss}.
As in the comoving frame the stationary sliding compound drop is of fixed shape, the contributions to the total energy $\mathcal{F}$ of capillary and wetting energy do not change in time, i.e., $\mathrm{d}\mathcal{F}_\mathrm{h}/\mathrm{d}t = 0$ (cf.~Eq.~\eqref{eq:F_horizontal}). Therefore, the only changing contribution is the potential energy in Eq.~\eqref{eq:ff2}. In consequence, it determines the total dissipation.\footnote{Note that even the hydrostatic contribution -- if it were considered -- would not contribute.}
In the nondimensionalized form (section~\ref{subsec:nondim-parameters-setup}) we obtain an expression for the total dissipation of stationary sliding compound drops $D = -\mathrm{d}\mathcal{F}_\mathrm{G}/\mathrm{d}t = -\beta\vel V$, where $V = V_1 + V_2$ is the total volume of the compound drop. In other words, at fixed volume, $D$  is proportional to $\beta\vel$, where $\vel$ itself depends on $\beta$, see Fig.~\ref{fig:plot_beta_variation}(a). Therefore, we are able to match the total dissipation of compound drops in the two configurations by simply adjusting $\beta$.

In particular, the first row of Fig.~\ref{fig:plot_Dissipation_cd} (panels (a) to (c)) shows the height-integrated spatially resolved dissipation for each liquid (dashed colored lines) as well as their sum (solid black lines) while the second row (Figs.~\ref{fig:plot_Dissipation_cd}(d)-(f)) gives the corresponding interface height profiles together with the streamlines and the color-coded magnitude of the velocity fields in the comoving frame. One clearly notices that in all cases the dissipation is mostly located in the three contact line regions at the substrate where pronounced peaks are visible. However, also the Neumann region strongly influences the flow field and the spatial distribution of dissipation. The {occurrence of peaks} in dissipation in the Young regions is in line with expectations from similar analyses for sliding drops of a single liquid, see Refs.~\cite{EWGT2016prf,MoVS2011el} and the one-drop limiting cases considered in appendix~\ref{sec:app_single_droplets}.

Note that although the sum of the dissipation in the two liquids is positive at any position along the substrate (solid black line in Figs.~\ref{fig:plot_Dissipation_cd}(a)-(c)), it can locally be positive in one liquid and negative in the other one. This occurs to a small extent when one liquid coexists with an adsorption layer of the other one. More importantly, it also occurs in the region between the three-phase contacts of the two liquids with the substrate and with the ambient medium. This indicates that potential energy 'set free' in the liquid that shows the negative dissipation is transferred into the other liquid where it is dissipated. Also this transfer is strongest in the Young regions.

Now we come back to the question why the {2-1 drop} slides at identical $\beta$ faster than the {1-2 drop} (Fig.~\ref{fig:plot_beta_variation}): Comparing the spatial structure of the dissipation for the {2-1 drop}  (Figs.~\ref{fig:plot_Dissipation_cd}(a) and~(c)) and the {1-2 drop} (Fig.~\ref{fig:plot_Dissipation_cd}(b)), we note that in both cases the dissipation is maximal in the Young region where substrate, liquid~1 and gas meet. This is independent of its location, be it at the front ({2-1 drop}) or at the back ({1-2 drop}) of the compound drop. The second largest peak occurs within liquid~1 in the liquid-liquid-substrate contact region. The main difference between the two configurations at identical $\beta$ (Figs.~\ref{fig:plot_Dissipation_cd}(a) and~(b)) is the much larger dissipation in the leading Young region in the {2-1 drop} than in the trailing Young region in the {1-2 drop}. This is consistent with the larger sliding velocity of the former. However, to ensure this is not just an effect of the total dissipation we also compare the dissipation profiles at approximately identical total dissipation (Figs.~\ref{fig:plot_Dissipation_cd}(b) and~(c)). Although the differences in peak heights are smaller, the overall features of the peaks still hold, and the 1-2~drop is still slower than the 2-1~drop although the latter slides down the smaller inclination.

{These observations} can be consistently explained considering the behavior of the (dynamic) Young angles: As liquid~1 has the smaller equilibrium contact angle at the employed $\VolVer=\VisVer=1$ it dominates the dissipation and roughly determines the velocity of the compound drop. The difference between 2-1 and 1-2 configuration then results from the different behavior of advancing and receding dynamic contact angle, {namely, that the} former increases with increasing velocity while the latter decreases. {This provides a physical explanation of the different behavior:} In the 1-2 configuration the dynamic Young angle $\vartheta_{13}$ is a decreasing receding angle thereby prolonging drop~1 while in the 2-1 configuration $\vartheta_{13}$ is an increasing advancing angle thereby shortening drop~1. This then ultimately results in the lower velocity of the drop in Fig.~\ref{fig:plot_Dissipation_cd}(e) than the one in Fig.~\ref{fig:plot_Dissipation_cd}(f).

Finally, we briefly discuss the spatial structure of the velocity fields in Figs.~\ref{fig:plot_Dissipation_cd}(d) and~(e). The pattern of convection rolls is structured by all three-phase contacts. In both configurations, liquid~2 features two lateral counter-rotating convection rolls, whereas in liquid~1 three convection rolls are visible - two of them are vertically staggered. This is consistent with the one-drop reference case in appendix~\ref{sec:app_single_droplets} where liquid~1 shows two vertically staggered convection rolls and liquid~2 shows only one. In comparison, the picture for the compound drop involves two additional lateral convection rolls related to the continuity of the liquid velocity across the liquid-liquid interface.

To summarize, we have found that at identical drop volumes and viscosities the velocity of the sliding compound drop is mainly determined by the liquid with the lower equilibrium contact angle as determined by the interface energies. The difference in velocities between the two configurations then ultimately results from the opposite behavior of the dynamic contact angles at the leading and trailing edge. Next, we briefly investigate how the overall picture changes when the volume ratio is varied.

\subsection{Influence of the volume ratio}
\label{subsec:var_of_volume}
\begin{figure}[!htb]
	\centering
	\includegraphics[width=\linewidth]{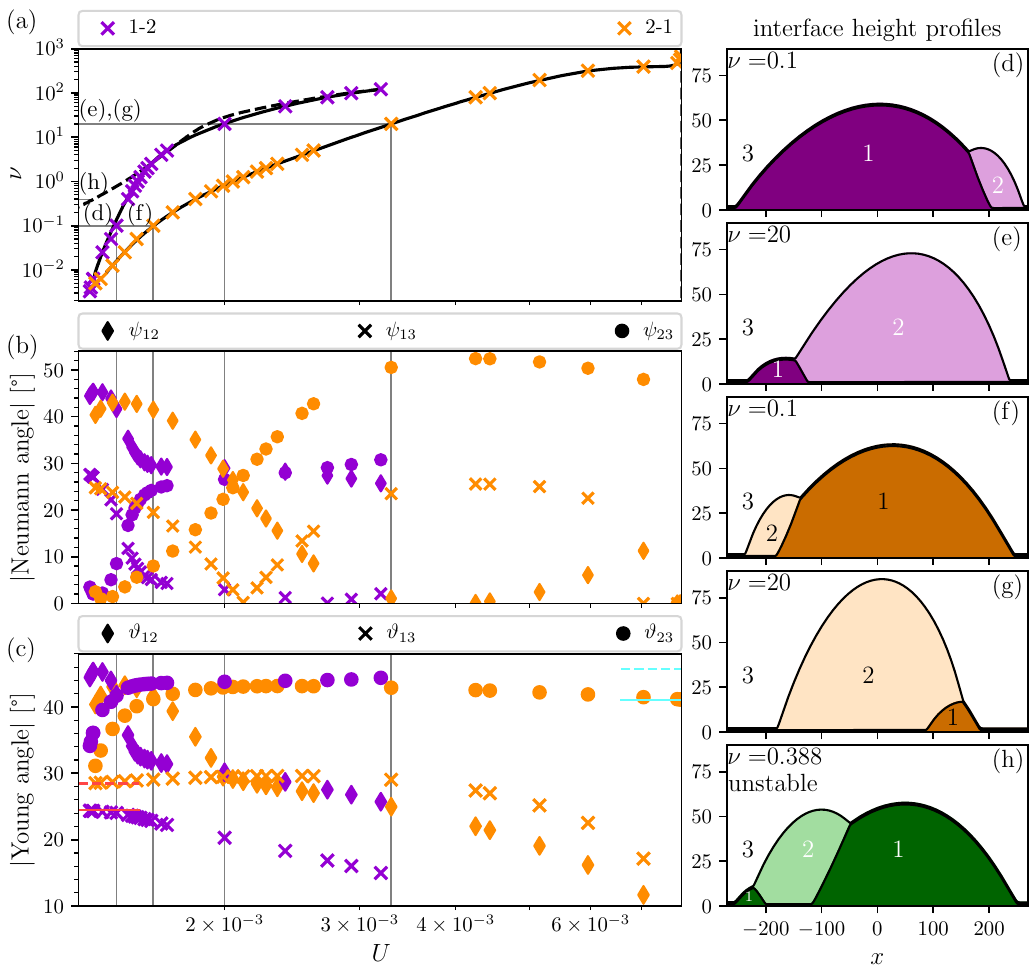}
	\caption{Dependence of the properties of stationary sliding compound drops on the volume ratio $\VolVer$ at fixed $\VisVer = 1$ and $\beta=10^{-5}$. Panel (a) shows the drop velocity $\vel$ (horizontal axis) as a function of $\VolVer$ (vertical axis)  in a log-log representation while panels~(b) and~(c) give the absolute values of the three Neumann angles and the three Young angles, respectively (see legends). Compound drops in 2-1 and 1-2 configuration are indicated by orange and purple coloring. Panels (d)-(g) and (h) present stable, and unstable states, respectively, at selected $\VolVer$ as indicated by thin solid horizontal lines in (a). The corresponding $\vel$-values are marked by vertical lines in (a)-(c). The left and right edges of panels (a)-(c) are chosen to the one-drop velocities of liquid~1 ($\VolVer\to 0$) and liquid~2 ($\VolVer\to\infty$), respectively (see Appendix \ref{sec:app_single_droplets}). Advancing and receding contact angles from the one-drop limiting cases are given in panel (c) as thin colored dashed and solid horizontal lines, respectively. Remaining parameters, lines styles etc.\ are as in Fig.~\ref{fig:plot_beta_variation}.}
	\label{fig:plot_volume_variation}
\end{figure}

{The analysis in section~\ref{subsec:var_of_inclination} is focused on the influence of driving force represented by the inclination $\beta$ at fixed volume ratio $\VolVer=1$. However, beside the inclination also the volume ratio is an experimental parameter that can be well controlled. Therefore we next investigate how the drop behavior depends on $\VolVer$ (at fixed $\VisVer=1$ and $\beta=10^{-5}$). Note that while changing $\VolVer$ we fix the total volume $V_\text{tot} = V_1 + V_2 = 2\times10^{4}$. This ensures that changes in the velocity $\vel$ do reflect changes in the drop configuration and are not dominated by changes in the total mass.}

{Again combining results obtained from numerical path continuation (solid and dashed lines) and time simulations (cross symbols), Fig.~\ref{fig:plot_volume_variation}(a) provides the dependence of drop velocity $\vel$ on $\VolVer$ in log-log representation. Also here, the 2-1~drop slides faster than the 1-2~drop} by up to approximately a factor two, while both velocities increase monotonically with increasing $\VolVer$. {At small and large $\VolVer$ the properties of the drops converge to the ones} of single-liquid drops of liquid~1 and liquid~2, respectively (appendix~\ref{sec:app_single_droplets}). The limiting values of the contact angles are marked in Figs.~\ref{fig:plot_volume_variation}(c) by thin horizontal colored lines (red and blue for a drop of liquid~1 and liquid~2, respectively).

{It is notable that for} $\VolVer\to 0$ both configurations approach the same velocity, while for $\VolVer\to\infty$ only the 2-1~drop approaches the proper limit. {In contrast, the} branch of stable 1-2~drops ceases to exist at a saddle-node bifurcation at $({\VolVer}_{\mathrm{c}12}, \vel_{\mathrm{c}12}) \approx (121.6, 3.5\times 10^{-3})$ {where it annihilates with a branch of unstable drop states. They correspond to a more complex compound drop in 1-2-1 configuration, i.e., a drop of liquid~1 is sandwiched by two drops of liquid~2 (see Fig.~\ref{fig:plot_volume_variation}(f)).} 

{To investigate the behavior beyond ${\VolVer}_{\mathrm{c}12}$, we use an equilibrated 1-2~drop with $\VolVer>{\VolVer}_{\mathrm{c}12}$ on a horizontal substrate ($\beta = 0$) to initiate a time simulation at the considered inclination $\beta=10^{-5}$. This shows that the already small trailing drop of liquid~1} 'dissolves' into the absorption layer. Due to the periodic boundary conditions {it gets later collected at the front of drop~2 where it forms a wetting ridge, i.e., the resulting configuration is a 2-1~drop.} The Young and Neumann angles (Figs. \ref{fig:plot_volume_variation}(b) and (c)) show a linear trend {as ${\VolVer}_{\mathrm{c}12}$ is approached}. {Note that a stationary sliding 2-1~drop is stable for all investigated $\VolVer$.}

Changes in the Neumann angles (Fig.~\ref{fig:plot_volume_variation}(b)) originate from the rotation of the Neumann region as $\VolVer$ is changed (see drop profiles in Figs.~\ref{fig:plot_volume_variation}(d)-(g)). In particular, all angles except of $\psi_{12}$ in the 1-2~configuration change their sign {once}. This results from a rotating of the Neumann region by at least 90 degrees that occurs for both configurations. Especially, {when approaching the limiting cases, the angles change nonlinearly because mesoscopic effects dominate when one of the drops becomes very small.}\footnote{As above we emphasize, that the Neumann residues remain on a small but finite constant level while all individual Neumann angles change. However, when approaching the two limiting cases these residues strongly increase. {There, the measure becomes less trustworthy because the Neumann angles are not well defined anymore (see rightmost orange data points in Fig.~\ref{fig:plot_volume_variation}(b)).}}
Similar observations can be made for the Young angles (Fig.~\ref{fig:plot_volume_variation}(c)): Over the course of the investigated $\VolVer$-range, only the receding angle $\vartheta_{13}$ and advancing angle $\vartheta_{23}$ change with increasing velocity as one would naively expect, {i.e., decrease and increase with increasing velocity, respectively. The other angles change nonmonotonically with velocity, especially for small $\VolVer$. As in the case of the Neumann angles, the reason lies in stronger mesoscopic effects and in the redistribution of mass and thereby dissipation between the two liquids that results from changing the volume ratio at fixed total volume.}

In summary, when varying the inclination parameter, the volume ratio, and the viscosity ratio {(see appendix~\ref{sec:app_var_of_viscosity})}, the velocities of stable stationary compound drops decrease or increase monotonically. 
Interestingly, the velocities of compound drops in the 2-1~configuration are for the investigated parameter ranges always larger than for drops in 1-2~configuration. {Considering the dissipation, we have shown that this can be explained as an effect resulting from interface energies that defined the smallest Young angle.}
{A remarkable observation for all studied parameter dependencies is that there are finite existence ranges for the two mainly considered configurations. Their existence range is limited by critical parameter values defined by saddle-node bifurcations. Beyond these critical points, either a configuration change toward another stationary sliding drop or a time-periodic drop splitting-fusion cycles emerge. The next section briefly analyzes the latter.}

\section{Transients and configuration changes}
\label{sec:transients}

\begin{figure}[t]
	\centering
	\includegraphics[width=\linewidth]{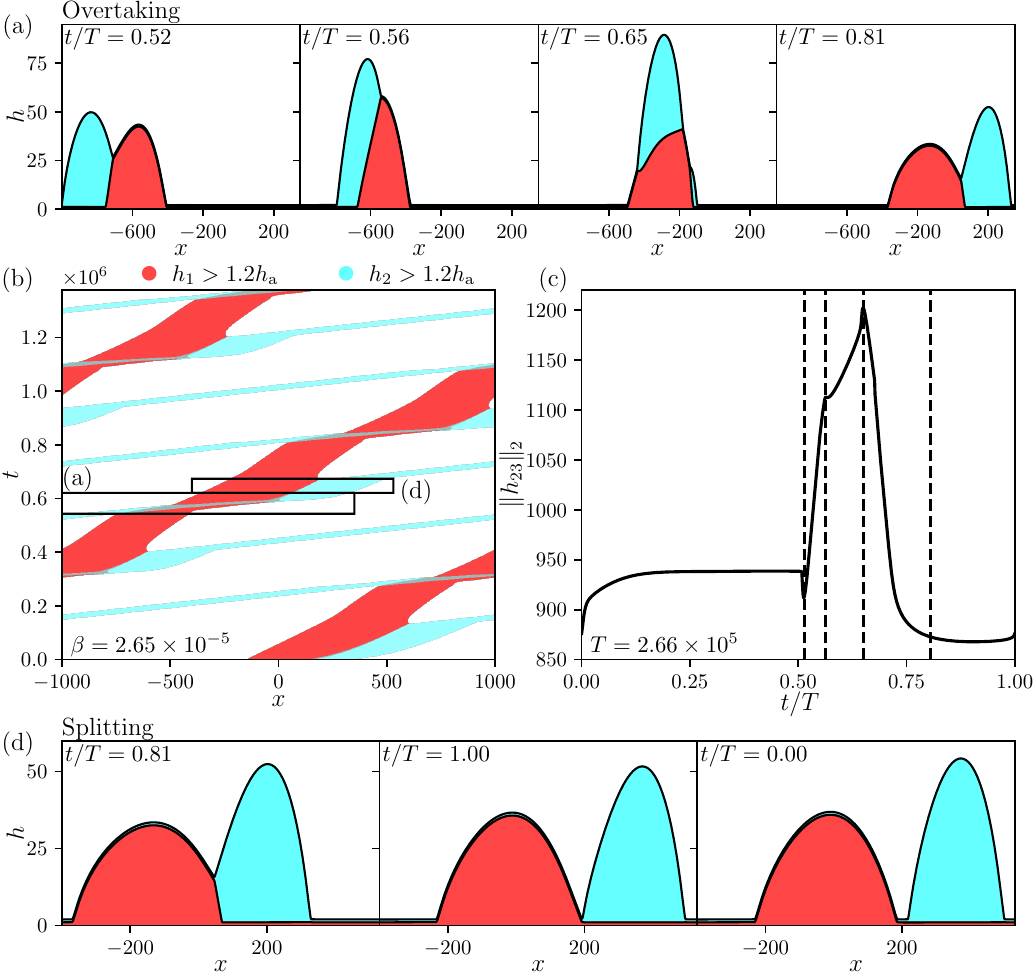}
	\caption{Dynamics of sliding compound drops at large inclination $\beta = 2.65\times 10^{-5}$. The space-time plot in panel~(b) provides an overview of the time-periodic process of fusing, overtaking and splitting of the  two drops of different liquids. The plot gives a top view of the distribution of the two liquids while only thicknesses above the threshold value of 1.2 contribute. The rows of snapshots show phases of (a) the overtaking and (d) the splitting dynamics. The selected time spans are marked with boxes in panel (b). Panel (c) shows the $\text{L}_2$-norm $\left\lVert\hTT\right\rVert_2$ over one period $T$ with vertical dashed lines corresponding to the times of the snapshots in (a). The domain size is $L=2000$ and the remaining parameters are the same as in Fig.~\ref{fig:plot_beta_variation}. {Also consider the corresponding movie~2 in the Supplemental Material.}}
	\label{fig:plot_Space-Time-Splitting}
\end{figure}
      
At large inclination parameters $\beta$ no stationary sliding compound drops exist {beyond the second saddle-node bifurcaton} (Fig.~\ref{fig:plot_beta_variation}). Instead, due to the used periodic boundary conditions interesting {time-periodic dynamics emerge}: The compound drop undergoes a periodic sequence of splitting, separated sliding, fusing and overtaking of {the two individual drops of the two liquids.} In an experiment on a long incline this would correspond to interacting arrays of periodically deposited drops of liquid~1 and liquid~2 similar to Ref.~\cite{PoFL2001prl} for sliding drops of a single liquid.

A typical example is presented in Fig.~\ref{fig:plot_Space-Time-Splitting}, %
{for a corresponding video see movie~2 in the Supplemental Material.} %
Fig~\ref{fig:plot_Space-Time-Splitting}(b) shows a space-time plot of several {periods $T$} after initial transients have passed. Starting at $t=0$ the compound drop first travels in slowly deforming 1-2 configuration till it splits into two separate droplets, see {the three} snapshots for this phase in Fig.~\ref{fig:plot_Space-Time-Splitting}(d). The splitting results from the faster velocity of the center of mass (CoM) of liquid~2 as compared to the CoM of liquid~1 (see Appendix~\ref{sec:app_single_droplets}). During the splitting process, the Neumann region approaches the substrate until, in passing, a four-phase contact region forms (second image of Fig.~\ref{fig:plot_Space-Time-Splitting}(d)). Afterwards, the drops slide individually with different velocities, i.e., the leading drop~2 moves increasingly ahead of the trailing drop~1.
Due to the periodic boundary conditions, the drop of liquid~2 closes in on the drop of liquid~1 until they fuse again. Then, they temporarily form a compound drop in 2-1 configuration. However, the trailing drop passes over the leading one such that a change occurs from a 2-1 to a 1-2 configuration, see {the four} snapshots in Fig.~\ref{fig:plot_Space-Time-Splitting}(a).

{The entire process} repeats periodically, for $ \beta = 2.65\times 10^{-5} $ with a period of $T=2.66\times10^5$. A compact overview of the phases of a cycle is provided in Fig.~\ref{fig:plot_Space-Time-Splitting}(c) that shows the L$_2$-norm $\left\lVert\hTT\right\rVert_2$ defined in section~\ref{subsec:nondim-parameters-setup} over one period. Starting from the left, the individual phases (approximately separated by the vertical dashed lines that indicate the times of the snapshots in Fig.~\ref{fig:plot_Space-Time-Splitting}(a)) {correspond to (i) sliding of separate drops where the norm remains nearly constant as the two drops move  individually with different velocities but without changing their individual shapes (ii) fusion into compound drop in 2-1 configuration visible as a brief drop in the norm (iii, iv) overtaking accompanied by a strong increase in the norm which peaks when drop~2 is on top of drop~1, (v) sliding as slowly separating compound drop in 1-2 configuration that then split again into two separate drops. The increase of the norm reflects the details of the separation process.}

Note that the time-periodic behavior remains qualitatively similar {to the shown example when further increasing $\beta$. However, the period $T$} monotonically decreases as the mean drop velocities increase. For the investigated parameter range, the behavior of the compound drops during one splitting-fusing-overtaking cycle is always {truly time-periodic, i.e., we have not observed differences between individual periods that would indicate a period-doubling bifurcation. Such effects are} observed for sliding drops of a single liquid on a two-dimensional substrate \cite{EWGT2016prf}, related to experimentally visible changes in the sequence of satellite drops \cite{PoFL2001prl}.

{Note, however,} that for extreme parameter values one can obtain irregular dynamics (not shown), namely, when due to large inclination the compound drop is stretched such that its length approaches the domain size. Then, a \enquote{dynamic wetting transition} occurs, and {a two-layer film with} large-amplitude interface waves emerges. At even larger $\beta$ one may then approximately {describe the system by} two coupled Kuramoto-Sivashinsky equations, similar to the system studied in Ref.~\cite{Klia1999jfm}. This is not further pursued here.

{Another interesting question is how the time-periodic behavior ceases when decreasing $\beta$, i.e., when moving towards the saddle-node bifurcation that limits the stationary sliding drops at $\beta_{\mathrm{c}12}$  (Fig.~\ref{fig:plot_beta_variation}(a)). Time simulations for $\beta < \beta_{\mathrm{c}12}$ (not shown) show that after a possibly long transient the system converges to the stationary 1-2 drop, i.e., there is no multistability. Approaching $\beta_{\mathrm{c}12}$ from above, the time period gets larger as the slowly deforming 1-2 configuration during the cycle takes longer till the splitting occurs. Measuring the mean velocities of the two individual liquids over one period $T$ one finds that both slowly tend to the one of the 1-2~drop at $\beta_{\mathrm{c}12}$. This allows us to formulate the hypothesis that the time-periodic states emerge in a global bifurcation, namely, a saddle-node infinite-period (SNIPER) bifurcation \cite{Strogatz2015} as also observed in other hydrodynamic systems \cite{TuBa1988prl,ThKn2006njp,MMLP2013prsapes,KoTh2014n,GrSt2024sm}.}
\section{Conclusion}
\label{sec:conc}
We have investigated sliding compound drops consisting of two immiscible non-volatile partially wetting liquids on an inclined homogeneous smooth solid substrate.  In particular, we have considered the mesoscopic hydrodynamic model for sessile compound drops {presented in \cite{DiTh2025prf} and focused on the case of an inclined substrate taking the potential energy related to the downhill direction into account.} Note that the model in \cite{DiTh2025prf} itself amends, extends and unifies a number of previous mesoscale models for two-layer films, compound drops, and drops on liquid substrates \cite{BrMR1993l,PBMT2004pre,PBMT2005jcp,FiGo2005jcis,BaGS2005iecr,CrMa2009rmp,BCJP2013epje,PBJS2018sr,Kita2024jns} by incorporating an improved wetting energy that can cover the full spectrum of macroscopic parameters. These are related to the mesoscopic parameters via consistency relations such that the mesoscopic hydrodynamic model  (that exists in long-wave and full-curvature variants) correctly reflects the macroscopic  Laplace, Neumann, and Young laws through mesoscopic counterparts \cite{DiTh2025prf}. This implies that for sliding drops on an incline one can {now properly investigate} dynamic Neumann, and Young angles. 

Employing periodic boundary conditions for a one-dimensional substrate the potential energy contribution keeps the system permanently out of equilibrium, even though the model  can be written as a gradient dynamics on an energy functional {that is, however, not a Lyapunov functional as it is not bounded from below}. Using the full-curvature variant {of the model} we have then investigated the dynamic behavior of (i) a drop of liquid on a layer of another liquid that acts as an adaptive substrate, and (ii) of sliding compound drops that occur in different configurations. %
{Thereby, the particular aim of the performed parameter study has been the identification of qualitative transitions where at specific values of the considered control parameters (driving force, ratios of drop volumes and viscosities) discontinuous transitions between different drop configurations and time-periodic behavior occur. Our interest has focused on the underlying bifurcation structure and on the quantitative changes in the properties of the sliding drops in the different configurations. Basing our analysis on a combination of numerical time simulations and path continuation has allowed us to study both stable and unstable stationary compound drops and emerging time-periodic behavior.}

For the drops sliding on the adaptive substrate we have found that {the asymmetry of the sliding state depends nonmonotonically on substrate inclination.} This results from competing influences of capillary forces that themselves depend on the dynamic Neumann angles that differ at the front and back of the droplet, and friction forces due to shear stress that due to the parabolic velocity profiles strongly depends on the thickness of the liquid substrate. The nonmonotonic dependence of velocity on the thickness of the liquid substrate should in the future be investigated more in details. It occurs at very small substrate thicknesses and parallels experimental observations from drops sliding on polymer brush layers of nanometric thickness \cite{ZWLS2024am}. There, the velocity first increases before decreasing again reflecting the behavior in the inset of Fig.~\ref{fig:drop-on-layer} (for $\bar{h}_{12}$ from 3 to 6).

In a more extensive {parameter} study, we have then considered the dependence of configuration and velocity of stationary sliding compound drops on {mainly two control parameters, i.e., the inclination parameter and the ratio of drop volumina (the dependence on the viscosity ratio is considered in appendix~\ref{sec:app_var_of_viscosity}).} A general result is that stable {stationary} compound drops occur in two possible configurations, {referred to as 1-2~drop and 2-1~drop. Interestingly, for the studied parameter ranges the 2-1~drop is always faster than the 1-2~drop. This has been explained analyzing} the spatial distribution of dissipation along the substrate (noting that only potential energy is 'freed' in proportion to sliding speed and inclination, and fully determines dissipation). As a result, we have found that dissipation is stronger in the drop of smaller equilibrium contact angle, i.e., the steeper and faster drop is 'slaved' to the shallower slower one (at volume and viscosity ratio on unity).

{Analyzing the dependence of the dynamic Young and Neumann angles on velocity we have found that the} advancing and receding Young angles do not always show the naively expected behavior, i.e., that they increase and decrease, respectively, with increasing velocity. This results from the coupled influence of the three adapting fluid-fluid interfaces and friction forces whose direction depends on the details of the internal convection roll structure. We have further found that in large parts of the considered parameter ranges the Neumann angles might strongly change, but the Neumann residues (that show how well the Neumann laws are fulfilled) remain at a constant small level. This implies that the Neumann region merely rotates {in agreement with results for contact lines on soft and adaptive substrates \cite{LWBD2014jfm,GrHT2023sm}.}

\begin{figure}
  \centering
   \includegraphics[width=0.99\linewidth]{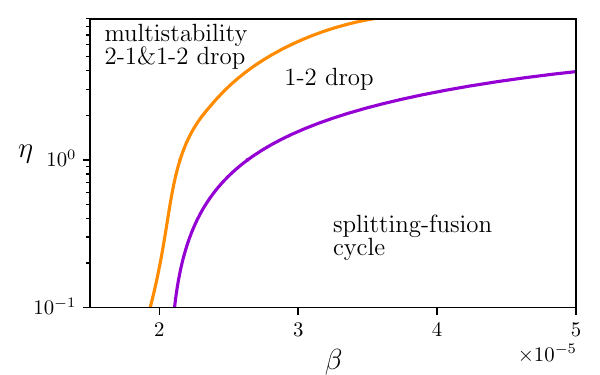}
   \caption{{Shown are the regions with qualitatively different behavior in the plane spanned by the inclination parameter $\beta$ and the viscosity ratio $\VisVer$. The borders between the regions where (i) the 2-1~drop and the 1-2~drop are both stable, (ii) only the 1-2 drop is stable, and (iii) splitting-fusion cycles occur are obtained via a two-parameter continuation of the saddle-node bifurcations of the branches of the 2-1 drop (orange) and of the 1-2 drop (magenta).}}
  \label{fig:plot_two_param_cont}
\end{figure}

{In general, the two configurations of stationary compound drops have finite existence ranges that are limited by saddle-node bifurcations occurring at some critical parameter value. Bringing a compound drop beyond such a critical value, different configuration changes take place either between 1-2~drop and 2-1~drop or between a stationary sliding drop and a cyclic splitting-fusion behavior. }
{Fig.~\ref{fig:plot_two_param_cont} summarizes these findings in the plane spanned by the inclination parameter $\beta$ and the viscosity ratio $\VisVer$. There, we indicate the three regions with qualitatively different behavior, namely, where (i) the 2-1~drop and the 1-2~drop are stable, (ii) only 1-2 drop is stable, and (iii) splitting-fusion cycles represent the stable dynamics. The orange and  the magenta line separating the regions correspond to the loci of the saddle-node bifurcation of the branches of 2-1~drop and 1-2~drop, respectively. They are obtained by a two-parameter continuation.}

The present study may be extended in a number of different directions: First, we emphasize that here all macroscopic interface energies, i.e., Hamaker constants and interface energies on the mesoscale, have been kept fixed. A {more} extensive parameter study might discover other drop configurations, bifurcation {structures}, and more intricate dynamics.
Second, an extension to two-dimensional substrates \cite{PBMT2006el,EWGT2016prf,DiTh2025prf} will allow to assess whether there exist further configurations of compounds drops and other types of dynamic transitions. Preliminary investigations indicate that a central overtaking exists (not shown). It will be particularly interesting whether the additional degree of freedom allows for more complex dynamics including long-time irregular behavior {as observed in \cite{EWGT2016prf} for drops of a single liquid.} Furthermore, one may incorporate additional external influences, e.g., {one may} investigate the influence of the hydrostatic part of the potential energy, assume the liquids are dielectric and incorporate an external electrical field \cite{LKBH2001jcp,VSKB2005l,ABCO2012sm}, {study the influence of inertia at large inclinations \cite{PoMa2023prf, Klia1999jfm}}, consider a nonisothermal setting \cite{PBMT2005jcp,NeSi2009prl,NeSi2021prf}, and study complex liquids, e.g., with surfactants \cite{MaCr2009sm,ThAP2016prf}.

\section*{Acknowledgments}

We acknowledge support by the Deutsche Forschungsgemeinschaft (DFG) via Grant No.\ TH781/12-1 and TH781/12-2 within Priority Program (SPP)~2171 \enquote{Dynamic Wetting of Flexible, Adaptive, and Switchable Surfaces}. We acknowledge preliminary work of Kevin Mitas on earlier models for two-layer systems published in his Ph.D. thesis. Furthermore, we acknowledge fruitful discussions with Daniel Greve, Christopher Henkel, and Florian Voss at the University of Münster, and with participants of the events organized by SPP 2171. Part of the calculations were performed on the HPC cluster PALMA II of the University of Münster, subsidized by the DFG via Grant No. INST 211/667-1.

\section*{Data availability statement}
The underlying data and {plot scripts required to generate the shown figures is publicly available at the
data repository \textit{zenodo} with the corresponding \textsc{doi}:\href{https://doi.org/10.5281/zenodo.19254325}{10.5281/zenodo.19254325} while the source code required to reproduce the time simulations can be found at the repository with the \textsc{doi}:\href{https://doi.org/10.5281/zenodo.13143462}{10.5281/zenodo.13143462}.}

\appendix

{
	\section{Nondimensional equations}
	\label{sec:app_nondim}
	This appendix provides details of the nondimensionalization employed in section~\ref{subsec:nondim-parameters-setup} and discusses the physical meaning of the nondimensional parameters. We start with the dimensional model~\ref{eq:thinFilm-interfaceHeights} and rescale the lateral and vertical coordinates via $\vec{r}=L\Tilde{\vec{r}}$ and $h_{ij} = L \Tilde{h}_{ij}$, respectively, where a tilde indicates a nondimensional quantity and the characteristic length scale is $L = h_{\mathrm{a1}}$. To scale time we choose the surface tension scaling (cf.~Ref.~\cite{Thie2007chapter}) $t=\tau\tilde{t}$ with $\tau=L\eta_1/\gamma_{12}$ implying a velocity scaling of $U = U_0 \Tilde{U}$ with $U_0 = \gamma_{12}/\eta_1$.
	Based on this scaling the nondimensional form of Eq.~\eqref{eq:thinFilm-interfaceHeights} is
	\begin{equation}
		\label{eq:thinFilm-interfaceHeights_nondimensional}
		\begin{aligned}	
			\partial_{\Tilde{t}} \Tilde{h}_{12} &= \nabla_{\Tilde{\vec{r}}} \cdot  \left(\Tilde{Q}_{11}\nabla_{\Tilde{\vec{r}}}\TildedFdhOT + \Tilde{Q}_{12}\nabla_{\Tilde{\vec{r}}}\TildedFdhTT\right),\\
			\partial_{\Tilde{t}} \Tilde{h}_{23} &= \nabla_{\Tilde{\vec{r}}} \cdot  \left(\Tilde{Q}_{21}\nabla_{\Tilde{\vec{r}}}\TildedFdhOT + \Tilde{Q}_{22}\nabla_{\Tilde{\vec{r}}}\TildedFdhTT\right),
		\end{aligned}
	\end{equation}
	with the nondimensional pressures
	\begin{align}
		\label{eq:dFdh12}
		\begin{split}	
			\TildedFdhOT &= -\frac{\nabla_{\Tilde{\vec{r}}}^2\Tilde{h}_{12}}{\Tilde{\xi}_{12}^3} + c_1\Tilde{f}_1^\prime - \Tilde{\xi}_{23}c_2\Tilde{f}_2^\prime + \rho_\mathrm{r} \Tilde{\beta} \Tilde{x},
		\end{split}\\ 
		\label{eq:dFdh23}
		\begin{split}	
			\TildedFdhTT &=  -\left(\frac{\gamTwoThr}{\gamOneTwo} + c_2\Tilde{f}_2\right) \frac{\nabla_{\Tilde{\vec{r}}}^2\Tilde{h}_{23}}{\Tilde{\xi}_{23}^3} + c_3\Tilde{f}_3^\prime + \Tilde{\beta} \Tilde{x}\\
			&~~~~ + \frac{1}{\Tilde{\xi}_{23}}\left(1+(\nabla_{\Tilde{\vec{r}}} \Tilde{h}_{12})\cdot(\nabla_{\Tilde{\vec{r}}}\Tilde{h}_{23})\right)c_2\Tilde{f}_2^\prime .
		\end{split}
	\end{align}
	Here, the $c_i = A_i/(\gamma_{12}L^{2})$ are dimensionless Hamaker constants and $\Tilde{\beta} = \beta L^{2}/\gamOneTwo=\rho_2 g L^{2}\sin\alpha/\gamOneTwo$ is the nondimensional inclination parameter that can be identified as a Bond number (cf.~Ref.~\cite{OrDB1997rmp}). Note that the mobility matrix as introduced in Eq.~\eqref{eq:Mobility_interfaceHeights} becomes $\tens{Q}=L^{3}/\eta_1 \tens{\Tilde{Q}}$ and the viscosity ratio $\eta=\eta_2/\eta_1$ remains its only free parameter. Overall, setting $\gamOneTwo = \eta_1 = h_{\mathrm{a1}} = 1$ as done in Sec.~\ref{subsec:nondim-parameters-setup} provides the same nondimensional evolution equations.
}

\section{Limiting cases of single droplets}
\label{sec:app_single_droplets}

\begin{figure}
	\centering
	\includegraphics[width=\linewidth]{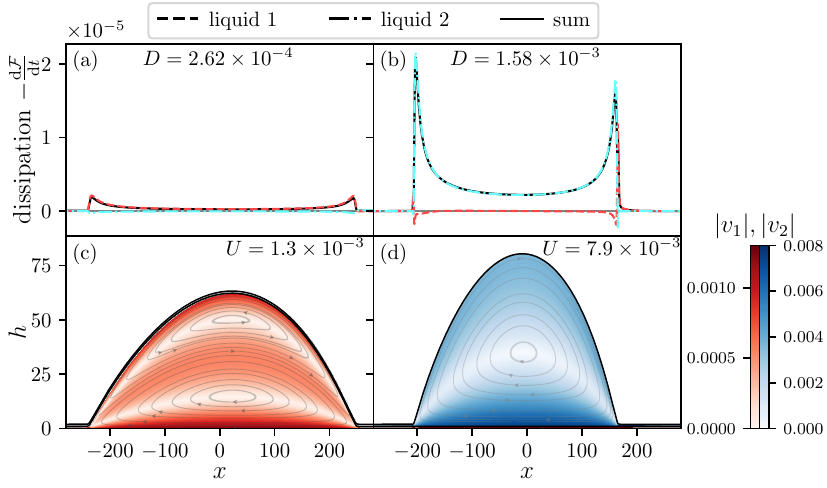}
	\caption{For stationary sliding single drops of liquid~1 (left) and liquid~2 (right) panels (a)-(b) give the laterally spatially resolved dissipation in the two liquids and their sum as well as the total dissipation $D$, and (c)-(d) the height profiles together with the streamlines and the color-coded magnitude of the velocity fields in the frame moving with velocity $\vel$ (that is indicated in the panels) at $\beta=10^{-5}$, $\VisVer=1$ and $V=2\times10^{4}$. The remaining parameters are as given in section~\ref{subsec:nondim-parameters-setup}.}
	\label{fig:plot_Dissipation_ref}
\end{figure}

{Here,} we briefly discuss two relevant limiting cases, namely, drops of a single liquid (either liquid~1 or liquid~2) that slide in the presence of only an adsorption layer of the other liquid ({for fixed $(\VisVer,\beta,V_i) = (1,10^{-5},2\times10^{4})$}), see Figs.~\ref{fig:plot_Dissipation_ref}(c) and (d). 
Receding and advancing dynamic contact angles as well as the sliding velocity $\vel$ are determined as described in section~\ref{subsec:nondim-parameters-setup}.

\begin{table}[!b]
	\caption{Equilibrium and advancing and receding dynamic contact angles of the reference cases. Parameters and numbering of columns correspond to Fig.~\ref{fig:plot_Dissipation_ref}.}
	\label{tab:erg_reference_angle}
	\begin{tabular}{@{}lcc@{}}
		\hline
		\hline
		& \multicolumn{2}{c}{\textbf{Reference case in Fig.~\ref{fig:plot_Dissipation_ref}}} \\ \hline 
		\textbf{Contact angle} & (c) $|\vartheta_{13}|$ & (d) $|\vartheta_{23}|$ \\ \hline
		Equilibrium analytically ($\beta=0$)& \SI{29,0}{\degree} & \SI{46.6}{\degree}  \\ \hline
		Equilibrium numerically ($\beta=0$)& \SI{26,7}{\degree} & \SI{43.6}{\degree}  \\ \hline
		Receding ($\beta=10^{-5}$)& \SI{24.5}{\degree} & \SI{41.1}{\degree} \\ \hline
		Advancing ($\beta=10^{-5}$)& \SI{28,5}{\degree} & \SI{45.7}{\degree}  \\ \hline
	\end{tabular}
\end{table}

Table~\ref{tab:erg_reference_angle} provides an overview of equilibrium and dynamic contact angles. For equilibrium Young angles the macroscopic values obtained from the macroscopic interface energies are compared to the mesoscopic values obtained for finite sized drops as slopes at the corresponding inflection points. The differences of about 8\% are a finite size effect and decrease when larger drops are considered with the mesoscopic model. 
As expected, the receding [advancing] angle is in both cases smaller [larger] than the numerically obtained equilibrium angle. Note that the drop of liquid~1 slides by about a factor of six slower than the drop of liquid~2 (cf.~Fig.~\ref{fig:plot_Dissipation_ref}(c) and (d)). Also this trend is expected as the drop of liquid~1 has a smaller equilibrium contact angle than the drop of liquid~2 $\vartheta_{13}<\vartheta_{23}$, i.e., the latter is taller and due to the strongly nonlinear mobility function $\sim h^3$ will slide much faster.

For further insights, Fig.~\ref{fig:plot_Dissipation_ref}(a) and (b) presents the spatially resolved dissipation (integrand of the Eq.~\eqref{eq:DissDef}). The liquid in the drop dominates the dissipation while the liquid in the adsorption layer gives a small negative contribution related to energy transfer between the liquids (see discussion in section~\ref{subsec:var_of_inclination}). In both cases, the dissipation is concentrated in the three-phase contact regions, in accordance with expectations \cite{EWGT2016prf}. 
The dissipation peak at the receding contact line is higher than at the advancing one which is also expected as the former contact angle is smaller than the latter.
As the drop of liquid~2 slides faster, more potential energy has to be dissipated per time, i.e., the total dissipation in the sliding drop of  liquid~2 has to be larger. Here, it is almost one order of magnitude larger than the dissipation in the drop of  liquid~1. 

Fig.~\ref{fig:plot_Dissipation_ref}(c) and (d) also give the streamlines as well as the color-coded magnitude of the velocity field (both in the comoving frame).  The drop of liquid~2 displayed in panel (d) features one convection roll with clockwise flux. In contrast, the drop of liquid~1 displayed in panel (c) shows two vertically staggered convection rolls. The lower one rotates clockwise and the upper one counterclockwise. The upper roll is a feature of the mesoscopic model and does not occur if the drop size is increased.

\section{Determination of dissipation}
\label{sec:app_diss}
The mesoscopic hydrodynamic (or thin-film) model corresponds to a gradient dynamics on an underlying energy functional that monotonically decreases in time \cite{Thie2018csa,DiTh2025prf}. The negative of the change in total energy directly corresponds to the total dissipation. However, the spatial distribution of dissipation may also be determined. Here, we consider the per liquid vertically integrated but laterally spatially resolved dissipation profiles. In particular  
$D_1(x,t)$ and $D_2(x,t)$ correspond to two dissipation {channels}, namely, in liquid~1 and~2, respectively. For a one-dimensional domain $\Omega$ one has
\begin{align}
	\begin{split}\label{eq:DissDef}
	D &= -\frac{\mathrm{d}\mathcal{F}}{\mathrm{d}t} = -\int_\Omega \sum_{\alpha=1}^2 \frac{\delta \mathcal{F}}{\delta h_\alpha}\partial_t h_\alpha \, \mathrm{d}x \\
	&= \int_\Omega \sum_{\alpha=1}^2 \frac{\delta \mathcal{F}}{\delta h_\alpha}(\partial_x j_\alpha) \, \mathrm{d}x \\ 
	&= \underbrace{\int_\Omega \left(\partial_x\frac{\delta \mathcal{F}}{\delta h_1}\right) \left[\sum_{\beta=1}^2 Q_{1\beta}\partial_x\frac{\delta\mathcal{F}}{\delta h_\beta}\right]\mathrm{d}x}_{D_1} \\
	& ~~~~+\underbrace{\int_\Omega \left(\partial_x\frac{\delta \mathcal{F}}{\delta h_2}\right) \left[\sum_{\beta=1}^2 Q_{2\beta}\partial_x\frac{\delta\mathcal{F}}{\delta h_\beta}\right]\mathrm{d}x}_{D_2}.
	\end{split}
\end{align}
where we use Eq.~\eqref{eq:thinFilm-interfaceHeights} in an alternative representation that uses layer thickness profiles $h_1$ and $h_2$ (and corresponding fluxes) instead of interface height profiles. The transformation between the different forms is discussed in some detail in \cite{DiTh2025prf}. The obtained expressions are time-independent for stationary sliding drops and provide the presented dissipation profiles.

{
\section{Influence of the viscosity ratio}
\label{sec:app_var_of_viscosity}
}

\begin{figure}
	\centering
	\includegraphics[width=\linewidth]{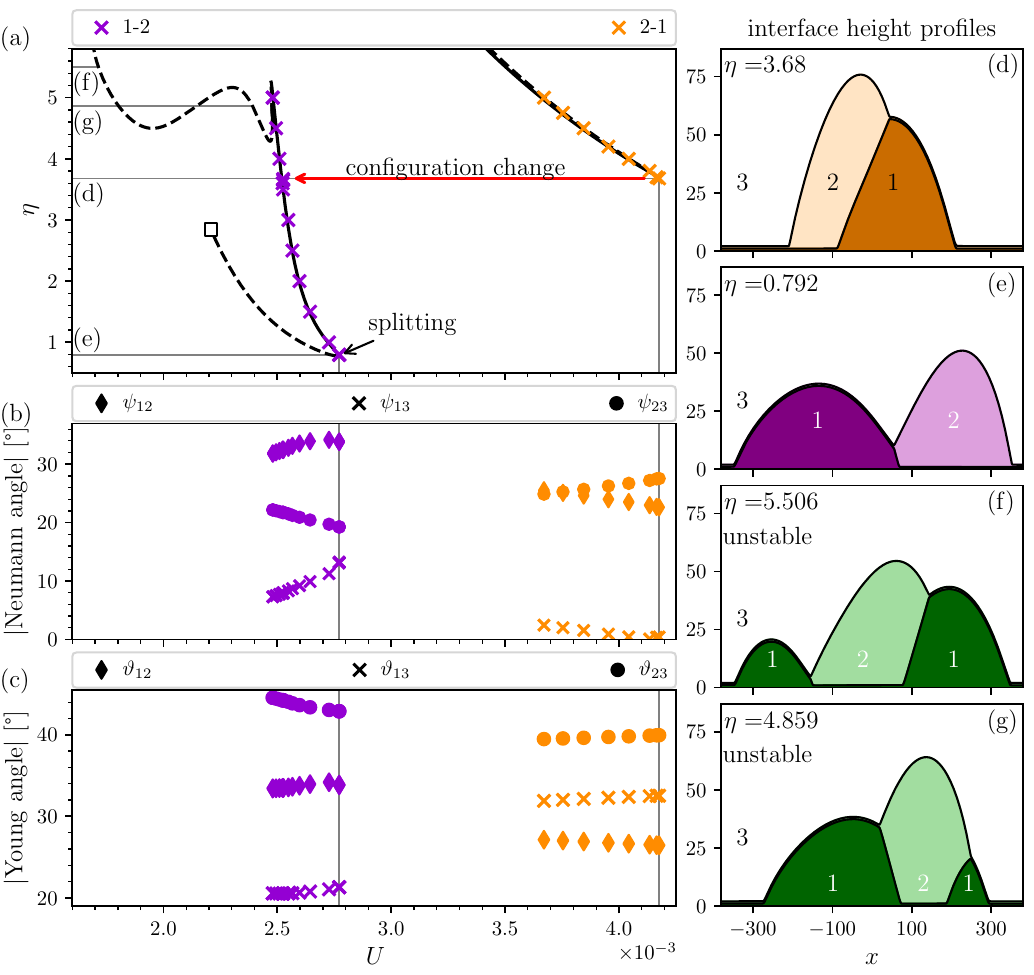}
  \caption{Dependence of the properties of stationary sliding compound drops on the viscosity ratio $\VisVer$ at fixed $\VolVer = 1$ and $\beta = 2.5 \times 10^{-5}$. Panel (a) shows the drop velocity $\vel$ (horizontal axis) as a function of $\VisVer$ (vertical axis) while panels~(b) and (c) give the absolute values of the three Neumann angles and the three Young angles, respectively (see legends). 
    At selected $\VisVer$, panels (d)-(e), and (f)-(g) present stable, and unstable states, respectively.
Remaining parameters, lines styles etc.\ are as in Fig.~\ref{fig:plot_beta_variation}.}
	\label{fig:plot_visc_variation}
\end{figure}

{Here,} we briefly investigate the influence of the viscosity ratio $\VisVer$ at fixed $\VolVer=1$ and $\beta = 2.5\times10^{-5}$. 
The results are displayed in Fig.~\ref{fig:plot_visc_variation}, where panel~(a) shows the dependence of the compound drop velocity $\vel$ on the viscosity ratio $\VisVer$ {with line styles and symbols as introduced in section~\ref{subsec:var_of_inclination}}.
Inspecting Fig.~\ref{fig:plot_visc_variation}(a) one notes that for stable compound drops in both configurations $\vel$ decreases with increasing $\VisVer$. {Again, the 2-1 drop slides}  nearly twice as fast as the  {1-2~drop}. 
This is in line with expectations, as for the given interface energies (and $\VolVer$) a single drop of liquid~2 indeed slides faster than an isolated drop of liquid~1 but we expect that the velocity difference becomes smaller with increasing viscosity ratio. 

For both configurations, we observe changes in the Neumann angles (Fig.~\ref{fig:plot_visc_variation}(b)) that again result from a rotation of the Neumann region. Considering the Young angles in Fig.~\ref{fig:plot_visc_variation}(c) {we note that besides the receding angle $\vartheta_{12}$ only the advancing angles $\vartheta_{13}$ and $\vartheta_{12}$ change with increasing velocity as one would naively expect. All other Young angles fall short of this expectation for the same reasons as discussed in section~\ref{subsec:var_of_volume}.}

Decreasing $\VisVer$, the {2-1~drop} ceases to exist at a critical value of ${\VisVer}_{\mathrm{c}21} \approx 3.68$, {corresponding to} $\vel_{\mathrm{c}21}\approx4.18\times10^{-3}$. Fig.~\ref{fig:plot_visc_variation}(d) shows the last stable {2-1~drop}. Decreasing $\VisVer$ below ${\VisVer}_{\mathrm{c}21}$ the liquid in drop~2 is moving through the adsorption layer over drop~1 dynamically forming a stable stationary sliding {1-2~drop} that moves at much lower velocity {(see movie~4 in the Supplemental Material)}.
Thus, this configuration change (indicated by an arrow in Fig.~\ref{fig:plot_visc_variation}(a)) differs from the overtaking process discussed in section~\ref{subsec:var_of_inclination} for $\beta > \beta_{\mathrm{c}21}$. We note that the Young and Neumann angles both show a linear trend {towards ${\VisVer}_{\mathrm{c}21}$} (see Figs.~\ref{fig:plot_visc_variation}(b) and (c)).

Decreasing the viscosity ratio even further, the {1-2~drop}  also ceases to exist in another saddle-node bifurcation at ${\VisVer}_{\mathrm{c}12}\approx 0.791$ ($\vel_{\mathrm{c}12} \approx 2.76\times10^{-3}$). Fig.~\ref{fig:plot_visc_variation}(e) shows the last stable {1-2~drop}. Below ${\VisVer}_{\mathrm{c}12}$ the compound drop splits into two distinct drops as described in section~\ref{sec:transients}. Following the unstable branch (that turns into the stable one at the saddle-node bifurcation) back to larger $\VisVer$ one finds unstable drops that first look similar to the one in Fig.~\ref{fig:plot_visc_variation}(e) and later correspond to two individual drops connected by a thin adsorption layer. They correspond to threshold states that have to be overcome to split a stable compound drop via a finite size perturbation (not shown). 
In the vicinity of this bifurcation, the Neumann angles $\psi_{13}$ and $\psi_{12}$ show nonlinear behavior while $\psi_{23}$ and the Young angles remain approximately linear (Figs.~\ref{fig:plot_visc_variation}(b) and~(c)). The  {unstable part of the branch is not followed further than the black square symbol in Fig.~\ref{fig:plot_visc_variation}(a)}.

Interestingly, the branch of stable {1-2~drops}  is as well limited toward large $\VisVer$ (Fig.~\ref{fig:plot_visc_variation}(a)). Namely, there exists another saddle-node bifurcation at ${\VisVer}_{\mathrm{c}12b} \approx 5.2$ ($\vel_{\mathrm{c}12b} \approx 2.5\times10^{-3}$) where the 1-2~drop ceases to exist. {The corresponding branch of unstable states again corresponds to drops in 1-2-1~configuration (\ref{fig:plot_visc_variation}(f) and (g)).} The unstable branch is nonmonotonic in $\VisVer$, there exists a range where four different states exist at identical $\VisVer$. 
If a stable {1-2~drop} is brought to a $\VisVer>{\VisVer}_{\mathrm{c}12b}$, time simulations indicate that liquid~1 then creeps through the adsorption layer underneath liquid~2 towards the leading edge of the compound drop. If $\VisVer$ is sufficiently far away from the saddle-node bifurcation, finally, a stable {2-1~drop}  is formed {(see movie~5 in the Supplemental Material)}. However, close but above ${\VisVer}_{\mathrm{c}12b}$ more involved time-periodic behavior can be found, namely, the compound drop can oscillate between both configurations. This forms an interesting starting point for a possible future detailed investigation of the bifurcation behavior.



\begin{thebibliography}{103}
\ifx \bisbn   \undefined \def \bisbn  #1{ISBN #1}\fi
\ifx \binits  \undefined \def \binits#1{#1}\fi
\ifx \bauthor  \undefined \def \bauthor#1{#1}\fi
\ifx \batitle  \undefined \def \batitle#1{#1}\fi
\ifx \bjtitle  \undefined \def \bjtitle#1{#1}\fi
\ifx \bvolume  \undefined \def \bvolume#1{\textbf{#1}}\fi
\ifx \byear  \undefined \def \byear#1{#1}\fi
\ifx \bissue  \undefined \def \bissue#1{#1}\fi
\ifx \bfpage  \undefined \def \bfpage#1{#1}\fi
\ifx \blpage  \undefined \def \blpage #1{#1}\fi
\ifx \burl  \undefined \def \burl#1{\textsf{#1}}\fi
\ifx \doiurl  \undefined \def \doiurl#1{\url{https://doi.org/#1}}\fi
\ifx \betal  \undefined \def \betal{\textit{et al.}}\fi
\ifx \binstitute  \undefined \def \binstitute#1{#1}\fi
\ifx \binstitutionaled  \undefined \def \binstitutionaled#1{#1}\fi
\ifx \bctitle  \undefined \def \bctitle#1{#1}\fi
\ifx \beditor  \undefined \def \beditor#1{#1}\fi
\ifx \bpublisher  \undefined \def \bpublisher#1{#1}\fi
\ifx \bbtitle  \undefined \def \bbtitle#1{#1}\fi
\ifx \bedition  \undefined \def \bedition#1{#1}\fi
\ifx \bseriesno  \undefined \def \bseriesno#1{#1}\fi
\ifx \blocation  \undefined \def \blocation#1{#1}\fi
\ifx \bsertitle  \undefined \def \bsertitle#1{#1}\fi
\ifx \bsnm \undefined \def \bsnm#1{#1}\fi
\ifx \bsuffix \undefined \def \bsuffix#1{#1}\fi
\ifx \bparticle \undefined \def \bparticle#1{#1}\fi
\ifx \barticle \undefined \def \barticle#1{#1}\fi
\bibcommenthead
\ifx \bconfdate \undefined \def \bconfdate #1{#1}\fi
\ifx \botherref \undefined \def \botherref #1{#1}\fi
\ifx \url \undefined \def \url#1{\textsf{#1}}\fi
\ifx \bchapter \undefined \def \bchapter#1{#1}\fi
\ifx \bbook \undefined \def \bbook#1{#1}\fi
\ifx \bcomment \undefined \def \bcomment#1{#1}\fi
\ifx \oauthor \undefined \def \oauthor#1{#1}\fi
\ifx \citeauthoryear \undefined \def \citeauthoryear#1{#1}\fi
\ifx \endbibitem  \undefined \def \endbibitem {}\fi
\ifx \bconflocation  \undefined \def \bconflocation#1{#1}\fi
\ifx \arxivurl  \undefined \def \arxivurl#1{\textsf{#1}}\fi
\csname PreBibitemsHook\endcsname

\bibitem[\protect\citeauthoryear{Young}{1805}]{Youn1805ptrs}
\begin{barticle}
\bauthor{\bsnm{Young}, \binits{T.}}:
\batitle{An essay on the cohesion of fluids}.
\bjtitle{Phil. Trans. R. Soc.}
\bvolume{95},
\bfpage{65}--\blpage{87}
(\byear{1805})
\doiurl{10.1098/rstl.1805.0005}
\end{barticle}
\endbibitem

\bibitem[\protect\citeauthoryear{de~Gennes}{1985}]{Genn1985rmp}
\begin{barticle}
\bauthor{\bsnm{Gennes}, \binits{P.G.}}:
\batitle{Wetting: {S}tatics and dynamics}.
\bjtitle{Rev. Mod. Phys.}
\bvolume{57},
\bfpage{827}--\blpage{863}
(\byear{1985})
\doiurl{10.1103/RevModPhys.57.827}
\end{barticle}
\endbibitem

\bibitem[\protect\citeauthoryear{de~Gennes et~al.}{2004}]{GennesBrochard-WyartQuere2004}
\begin{bbook}
\bauthor{\bsnm{Gennes}, \binits{P.-G.}},
\bauthor{\bsnm{Brochard-Wyart}, \binits{F.}},
\bauthor{\bsnm{Qu{\'e}r{\'e}}, \binits{D.}}:
\bbtitle{Capillarity and Wetting Phenomena: Drops, Bubbles, Pearls, Waves}.
\bpublisher{Springer},
\blocation{New York}
(\byear{2004}).
\doiurl{10.1007/978-0-387-21656-0}
\end{bbook}
\endbibitem

\bibitem[\protect\citeauthoryear{Bormashenko}{2009}]{Borm2009csaea}
\begin{barticle}
\bauthor{\bsnm{Bormashenko}, \binits{E.}}:
\batitle{Young, {B}oruvka-{N}eumann, {W}enzel and {C}assie-{B}axter equations as the transversality conditions for the variational problem of wetting}.
\bjtitle{Colloid Surf. A-Physicochem. Eng. Asp.}
\bvolume{345},
\bfpage{163}--\blpage{165}
(\byear{2009})
\doiurl{10.1016/j.colsurfa.2009.04.054}
\end{barticle}
\endbibitem

\bibitem[\protect\citeauthoryear{Podgorski et~al.}{2001}]{PoFL2001prl}
\begin{barticle}
\bauthor{\bsnm{Podgorski}, \binits{T.}},
\bauthor{\bsnm{Flesselles}, \binits{J.-M.}},
\bauthor{\bsnm{Limat}, \binits{L.}}:
\batitle{Corners, cusps, and pearls in running drops}.
\bjtitle{Phys. Rev. Lett.}
\bvolume{87},
\bfpage{036102}
(\byear{2001})
\doiurl{10.1103/PhysRevLett.87.036102}
\end{barticle}
\endbibitem

\bibitem[\protect\citeauthoryear{Le~Grand et~al.}{2005}]{LeDL2005jfm}
\begin{barticle}
\bauthor{\bsnm{Le~Grand}, \binits{N.}},
\bauthor{\bsnm{Daerr}, \binits{A.}},
\bauthor{\bsnm{Limat}, \binits{L.}}:
\batitle{Shape and motion of drops sliding down an inclined plane}.
\bjtitle{J. Fluid Mech.}
\bvolume{541},
\bfpage{293}--\blpage{315}
(\byear{2005})
\doiurl{10.1017/s0022112005006105}
\end{barticle}
\endbibitem

\bibitem[\protect\citeauthoryear{Snoeijer and Andreotti}{2013}]{SnAn2013arfm}
\begin{barticle}
\bauthor{\bsnm{Snoeijer}, \binits{J.H.}},
\bauthor{\bsnm{Andreotti}, \binits{B.}}:
\batitle{Moving contact lines: {S}cales, regimes, and dynamical transitions}.
\bjtitle{Annu. Rev. Fluid Mech.}
\bvolume{45},
\bfpage{269}--\blpage{292}
(\byear{2013})
\doiurl{10.1146/annurev-fluid-011212-140734}
\end{barticle}
\endbibitem

\bibitem[\protect\citeauthoryear{Engelnkemper et~al.}{2016}]{EWGT2016prf}
\begin{barticle}
\bauthor{\bsnm{Engelnkemper}, \binits{S.}},
\bauthor{\bsnm{Wilczek}, \binits{M.}},
\bauthor{\bsnm{Gurevich}, \binits{S.V.}},
\bauthor{\bsnm{Thiele}, \binits{U.}}:
\batitle{Morphological transitions of sliding drops - dynamics and bifurcations}.
\bjtitle{Phys. Rev. Fluids}
\bvolume{1},
\bfpage{073901}
(\byear{2016})
\doiurl{10.1103/PhysRevFluids.1.073901}
{\href{https://arxiv.org/abs/http://arxiv.org/abs/1607.05482}{{http://arxiv.org/abs/1607.05482}}}
\end{barticle}
\endbibitem

\bibitem[\protect\citeauthoryear{Fern{\'{a}}ndez-Toledano et~al.}{2019}]{FBLD2019jcis}
\begin{barticle}
\bauthor{\bsnm{Fern{\'{a}}ndez-Toledano}, \binits{J.-C.}},
\bauthor{\bsnm{Blake}, \binits{T.D.}},
\bauthor{\bsnm{Limat}, \binits{L.}},
\bauthor{\bsnm{De~Coninck}, \binits{J.}}:
\batitle{A molecular-dynamics study of sliding liquid nanodrops: {D}ynamic contact angles and the pearling transition}.
\bjtitle{J. Colloid Interface Sci.}
\bvolume{548},
\bfpage{66}--\blpage{76}
(\byear{2019})
\doiurl{10.1016/j.jcis.2019.03.094}
\end{barticle}
\endbibitem

\bibitem[\protect\citeauthoryear{Tsao et~al.}{2023}]{TsLT2023pf}
\begin{barticle}
\bauthor{\bsnm{Tsao}, \binits{Y.H.}},
\bauthor{\bsnm{Liao}, \binits{Y.C.}},
\bauthor{\bsnm{Tsao}, \binits{H.K.}}:
\batitle{Sliding motion of highly deformed droplets on smooth and rough surfaces: shape oscillation, chaotic breakage, corner shape, and pearling}.
\bjtitle{Phys. Fluids}
\bvolume{35},
\bfpage{122121}
(\byear{2023})
\doiurl{10.1063/5.0181630}
\end{barticle}
\endbibitem

\bibitem[\protect\citeauthoryear{Thiele}{}]{Thie2026epje}
\begin{botherref}
\oauthor{\bsnm{Thiele}, \binits{U.}}:
What is active wetting?
Eur. Phys. J. E
\textbf{49}(7),
56
\doiurl{10.1140/epje/s10189-026-00590-y}
\end{botherref}
\endbibitem

\bibitem[\protect\citeauthoryear{Kim et~al.}{2002}]{KiLK2002jcis}
\begin{barticle}
\bauthor{\bsnm{Kim}, \binits{H.Y.}},
\bauthor{\bsnm{Lee}, \binits{H.J.}},
\bauthor{\bsnm{Kang}, \binits{B.H.}}:
\batitle{Sliding of liquid drops down an inclined solid surface}.
\bjtitle{J. Colloid Interface Sci.}
\bvolume{247},
\bfpage{372}--\blpage{380}
(\byear{2002})
\doiurl{10.1006/jcis.2001.8156}
\end{barticle}
\endbibitem

\bibitem[\protect\citeauthoryear{Voinov}{1976}]{Voin1976fd}
\begin{barticle}
\bauthor{\bsnm{Voinov}, \binits{O.V.}}:
\batitle{Hydrodynamics of wetting}.
\bjtitle{Fluid Dyn.}
\bvolume{11},
\bfpage{714}--\blpage{721}
(\byear{1976})
\doiurl{10.1007/BF01012963}
\end{barticle}
\endbibitem

\bibitem[\protect\citeauthoryear{Hocking}{1983}]{Hock1983qjmam}
\begin{barticle}
\bauthor{\bsnm{Hocking}, \binits{L.M.}}:
\batitle{The spreading of a thin drop by gravity and capillarity}.
\bjtitle{Q. J. Mech. Appl. Math.}
\bvolume{36},
\bfpage{55}--\blpage{69}
(\byear{1983})
\doiurl{10.1093/qjmam/36.1.55}
\end{barticle}
\endbibitem

\bibitem[\protect\citeauthoryear{Cox}{1986}]{Cox1986jfm}
\begin{barticle}
\bauthor{\bsnm{Cox}, \binits{R.G.}}:
\batitle{The dynamics of the spreading of liquids on a solid surface. {P}art 1. viscous flow}.
\bjtitle{J. Fluid Mech.}
\bvolume{168},
\bfpage{169}--\blpage{194}
(\byear{1986})
\doiurl{10.1017/S0022112086000332}
\end{barticle}
\endbibitem

\bibitem[\protect\citeauthoryear{Mohammad~Karim}{2022}]{Moha2022jap}
\begin{botherref}
\oauthor{\bsnm{Mohammad~Karim}, \binits{A.}}:
A review of physics of moving contact line dynamics models and its applications in interfacial science.
J. Appl. Phys.
\textbf{132}
(2022)
\doiurl{10.1063/5.0102028}
\end{botherref}
\endbibitem

\bibitem[\protect\citeauthoryear{Peschka}{2018}]{Pesc2018pf}
\begin{barticle}
\bauthor{\bsnm{Peschka}, \binits{D.}}:
\batitle{Variational approach to dynamic contact angles for thin films}.
\bjtitle{Phys. Fluids}
\bvolume{30},
\bfpage{082115}
(\byear{2018})
\doiurl{10.1063/1.5040985}
\end{barticle}
\endbibitem

\bibitem[\protect\citeauthoryear{Etha et~al.}{2021}]{EDSD2021m}
\begin{barticle}
\bauthor{\bsnm{Etha}, \binits{S.A.}},
\bauthor{\bsnm{Desai}, \binits{P.R.}},
\bauthor{\bsnm{Sachar}, \binits{H.S.}},
\bauthor{\bsnm{Das}, \binits{S.}}:
\batitle{Wetting dynamics on solvophilic, soft, porous, and responsive surfaces}.
\bjtitle{Macromolecules}
\bvolume{54},
\bfpage{584}--\blpage{596}
(\byear{2021})
\doiurl{10.1021/acs.macromol.0c02234}
\end{barticle}
\endbibitem

\bibitem[\protect\citeauthoryear{Hartmann et~al.}{2024}]{HDGT2024l}
\begin{barticle}
\bauthor{\bsnm{Hartmann}, \binits{S.}},
\bauthor{\bsnm{Diekmann}, \binits{J.}},
\bauthor{\bsnm{Greve}, \binits{D.}},
\bauthor{\bsnm{Thiele}, \binits{U.}}:
\batitle{Drops on polymer brushes -- {A}dvances in thin-film modelling of adaptive substrates}.
\bjtitle{Langmuir}
\bvolume{40},
\bfpage{4001}--\blpage{4021}
(\byear{2024})
\doiurl{10.1021/acs.langmuir.3c03313}
\end{barticle}
\endbibitem

\bibitem[\protect\citeauthoryear{Zhou et~al.}{2024}]{ZWLS2024am}
\begin{barticle}
\bauthor{\bsnm{Zhou}, \binits{X.}},
\bauthor{\bsnm{Wang}, \binits{Y.}},
\bauthor{\bsnm{Li}, \binits{X.}},
\bauthor{\bsnm{Sudersan}, \binits{P.}},
\bauthor{\bsnm{Amann-Winkel}, \binits{K.}},
\bauthor{\bsnm{Koynov}, \binits{K.n.}},
\bauthor{\bsnm{Nagata}, \binits{Y.}},
\bauthor{\bsnm{Berger}, \binits{R.}},
\bauthor{\bsnm{Butt}, \binits{H.-J.}}:
\batitle{Thickness of nanoscale poly(dimethylsiloxane) layers determines the motion of sliding water drops}.
\bjtitle{Adv. Mater.}
\bvolume{18},
\bfpage{2311470}
(\byear{2024})
\doiurl{10.1002/adma.202311470}
\end{barticle}
\endbibitem

\bibitem[\protect\citeauthoryear{Karpitschka et~al.}{2016}]{KPLW2016pnasusa}
\begin{barticle}
\bauthor{\bsnm{Karpitschka}, \binits{S.}},
\bauthor{\bsnm{Pandey}, \binits{A.}},
\bauthor{\bsnm{Lubbers}, \binits{L.A.}},
\bauthor{\bsnm{Weijs}, \binits{J.H.}},
\bauthor{\bsnm{Botto}, \binits{L.}},
\bauthor{\bsnm{Das}, \binits{S.}},
\bauthor{\bsnm{Andreotti}, \binits{B.}},
\bauthor{\bsnm{Snoeijer}, \binits{J.H.}}:
\batitle{Liquid drops attract or repel by the inverted {C}heerios effect}.
\bjtitle{Proc. Natl. Acad. Sci. U. S. A.}
\bvolume{113},
\bfpage{7403}--\blpage{7407}
(\byear{2016})
\doiurl{10.1073/pnas.1601411113}
\end{barticle}
\endbibitem

\bibitem[\protect\citeauthoryear{Andreotti and Snoeijer}{2020}]{AnSn2020arfm}
\begin{barticle}
\bauthor{\bsnm{Andreotti}, \binits{B.}},
\bauthor{\bsnm{Snoeijer}, \binits{J.H.}}:
\batitle{Statics and dynamics of soft wetting}.
\bjtitle{Annu. Rev. Fluid Mech.}
\bvolume{52},
\bfpage{285}--\blpage{308}
(\byear{2020})
\doiurl{10.1146/annurev-fluid-010719-060147}
\end{barticle}
\endbibitem

\bibitem[\protect\citeauthoryear{Guan et~al.}{2015}]{GWXM2015l}
\begin{barticle}
\bauthor{\bsnm{Guan}, \binits{J.H.}},
\bauthor{\bsnm{Wells}, \binits{G.G.}},
\bauthor{\bsnm{Xu}, \binits{B.}},
\bauthor{\bsnm{McHale}, \binits{G.}},
\bauthor{\bsnm{Wood}, \binits{D.}},
\bauthor{\bsnm{Martin}, \binits{J.}},
\bauthor{\bsnm{Stuart-Cole}, \binits{S.}}:
\batitle{Evaporation of sessile droplets on slippery liquid-infused porous surfaces (slips)}.
\bjtitle{Langmuir}
\bvolume{31},
\bfpage{11781}--\blpage{11789}
(\byear{2015})
\doiurl{10.1021/acs.langmuir.5b03240}
\end{barticle}
\endbibitem

\bibitem[\protect\citeauthoryear{Semprebon et~al.}{2017}]{SeMK2017sm}
\begin{barticle}
\bauthor{\bsnm{Semprebon}, \binits{C.}},
\bauthor{\bsnm{McHale}, \binits{G.}},
\bauthor{\bsnm{Kusumaatmaja}, \binits{H.}}:
\batitle{Apparent contact angle and contact angle hysteresis on liquid infused surfaces}.
\bjtitle{Soft Matter}
\bvolume{13},
\bfpage{101}--\blpage{110}
(\byear{2017})
\doiurl{10.1039/c6sm00920d}
\end{barticle}
\endbibitem

\bibitem[\protect\citeauthoryear{Tress et~al.}{2017}]{TKPS2017sm}
\begin{barticle}
\bauthor{\bsnm{Tress}, \binits{M.}},
\bauthor{\bsnm{Karpitschka}, \binits{S.}},
\bauthor{\bsnm{Papadopoulos}, \binits{P.}},
\bauthor{\bsnm{Snoeijer}, \binits{J.H.}},
\bauthor{\bsnm{Vollmer}, \binits{D.}},
\bauthor{\bsnm{Butt}, \binits{H.-J.}}:
\batitle{Shape of a sessile drop on a flat surface covered with a liquid film}.
\bjtitle{Soft Matter}
\bvolume{13},
\bfpage{3760}--\blpage{3767}
(\byear{2017})
\doiurl{10.1039/c7sm00437k}
\end{barticle}
\endbibitem

\bibitem[\protect\citeauthoryear{Keiser et~al.}{2017}]{KKCQ2017sm}
\begin{barticle}
\bauthor{\bsnm{Keiser}, \binits{A.}},
\bauthor{\bsnm{Keiser}, \binits{L.}},
\bauthor{\bsnm{Clanet}, \binits{C.}},
\bauthor{\bsnm{Qu{\'e}r{\'e}}, \binits{D.}}:
\batitle{Drop friction on liquid-infused materials}.
\bjtitle{Soft Matter}
\bvolume{13},
\bfpage{6981}--\blpage{6987}
(\byear{2017})
\doiurl{10.1039/c7sm01226h}
\end{barticle}
\endbibitem

\bibitem[\protect\citeauthoryear{Luo et~al.}{2017}]{LGGM2017pra}
\begin{barticle}
\bauthor{\bsnm{Luo}, \binits{J.T.}},
\bauthor{\bsnm{Geraldi}, \binits{N.R.}},
\bauthor{\bsnm{Guan}, \binits{J.H.}},
\bauthor{\bsnm{McHale}, \binits{G.}},
\bauthor{\bsnm{Wells}, \binits{G.G.}},
\bauthor{\bsnm{Fu}, \binits{Y.Q.}}:
\batitle{Slippery liquid-infused porous surfaces and droplet transportation by surface acoustic waves}.
\bjtitle{Phys. Rev. Appl.}
\bvolume{7},
\bfpage{014017}
(\byear{2017})
\doiurl{10.1103/PhysRevApplied.7.014017}
\end{barticle}
\endbibitem

\bibitem[\protect\citeauthoryear{McHale et~al.}{2025}]{MJBW2025prl}
\begin{botherref}
\oauthor{\bsnm{McHale}, \binits{G.}},
\oauthor{\bsnm{Janahi}, \binits{S.}},
\oauthor{\bsnm{Barrio-Zhang}, \binits{H.}},
\oauthor{\bsnm{Wang}, \binits{Y.}},
\oauthor{\bsnm{Chen}, \binits{J.}},
\oauthor{\bsnm{Wells}, \binits{G.G.}},
\oauthor{\bsnm{Ledesma-Aguilar}, \binits{R.}}:
Adhesive forces in droplet kinetic friction on liquidlike surfaces.
Phys. Rev. Lett.
\textbf{135}
(2025)
\doiurl{10.1103/ljbc-k193}
\end{botherref}
\endbibitem

\bibitem[\protect\citeauthoryear{Li et~al.}{2022}]{LBSB2022np}
\begin{barticle}
\bauthor{\bsnm{Li}, \binits{X.}},
\bauthor{\bsnm{Bista}, \binits{P.}},
\bauthor{\bsnm{Stetten}, \binits{A.Z.}},
\bauthor{\bsnm{Bonart}, \binits{H.}},
\bauthor{\bsnm{Sch{\"u}r}, \binits{M.T.}},
\bauthor{\bsnm{Hardt}, \binits{S.}},
\bauthor{\bsnm{Bodziony}, \binits{F.}},
\bauthor{\bsnm{Marschall}, \binits{H.}},
\bauthor{\bsnm{Saal}, \binits{A.}},
\bauthor{\bsnm{Deng}, \binits{X.}},
\bauthor{\bsnm{Berger}, \binits{R.}},
\bauthor{\bsnm{Weber}, \binits{S.A.L.}},
\bauthor{\bsnm{Butt}, \binits{H.-J.}}:
\batitle{Spontaneous charging affects the motion of sliding drops}.
\bjtitle{Nat. Phys.}
\bvolume{18},
\bfpage{713}--\blpage{719}
(\byear{2022})
\doiurl{10.1038/s41567-022-01563-6}
\end{barticle}
\endbibitem

\bibitem[\protect\citeauthoryear{Brochard-Wyart et~al.}{1993}]{BrMR1993l}
\begin{barticle}
\bauthor{\bsnm{Brochard-Wyart}, \binits{F.}},
\bauthor{\bsnm{Martin}, \binits{P.}},
\bauthor{\bsnm{Redon}, \binits{C.}}:
\batitle{Liquid/liquid dewetting}.
\bjtitle{Langmuir}
\bvolume{9},
\bfpage{3682}--\blpage{3690}
(\byear{1993})
\doiurl{10.1021/la00036a053}
\end{barticle}
\endbibitem

\bibitem[\protect\citeauthoryear{Pototsky et~al.}{2004}]{PBMT2004pre}
\begin{barticle}
\bauthor{\bsnm{Pototsky}, \binits{A.}},
\bauthor{\bsnm{Bestehorn}, \binits{M.}},
\bauthor{\bsnm{Merkt}, \binits{D.}},
\bauthor{\bsnm{Thiele}, \binits{U.}}:
\batitle{Alternative pathways of dewetting for a thin liquid two-layer film}.
\bjtitle{Phys. Rev. E}
\bvolume{70},
\bfpage{025201}
(\byear{2004})
\doiurl{10.1103/PhysRevE.70.025201}
\end{barticle}
\endbibitem

\bibitem[\protect\citeauthoryear{Pototsky et~al.}{2005}]{PBMT2005jcp}
\begin{barticle}
\bauthor{\bsnm{Pototsky}, \binits{A.}},
\bauthor{\bsnm{Bestehorn}, \binits{M.}},
\bauthor{\bsnm{Merkt}, \binits{D.}},
\bauthor{\bsnm{Thiele}, \binits{U.}}:
\batitle{Morphology changes in the evolution of liquid two-layer films}.
\bjtitle{J. Chem. Phys.}
\bvolume{122},
\bfpage{224711}
(\byear{2005})
\doiurl{10.1063/1.1927512}
\end{barticle}
\endbibitem

\bibitem[\protect\citeauthoryear{Fisher and Golovin}{2005}]{FiGo2005jcis}
\begin{barticle}
\bauthor{\bsnm{Fisher}, \binits{L.S.}},
\bauthor{\bsnm{Golovin}, \binits{A.A.}}:
\batitle{Nonlinear stability analysis of a two-layer thin liquid film: {D}ewetting and autophobic behavior.}
\bjtitle{J. Colloid Interface Sci.}
\bvolume{291},
\bfpage{515}--\blpage{528}
(\byear{2005})
\doiurl{10.1016/j.jcis.2005.05.024}
\end{barticle}
\endbibitem

\bibitem[\protect\citeauthoryear{Bandyopadhyay et~al.}{2005}]{BaGS2005iecr}
\begin{barticle}
\bauthor{\bsnm{Bandyopadhyay}, \binits{D.}},
\bauthor{\bsnm{Gulabani}, \binits{R.}},
\bauthor{\bsnm{Sharma}, \binits{A.}}:
\batitle{Instability and dynamics of thin liquid bilayers}.
\bjtitle{Ind. Eng. Chem. Res.}
\bvolume{44},
\bfpage{1259}--\blpage{1272}
(\byear{2005})
\doiurl{10.1021/ie049640r}
\end{barticle}
\endbibitem

\bibitem[\protect\citeauthoryear{Pototsky et~al.}{2006}]{PBMT2006el}
\begin{barticle}
\bauthor{\bsnm{Pototsky}, \binits{A.}},
\bauthor{\bsnm{Bestehorn}, \binits{M.}},
\bauthor{\bsnm{Merkt}, \binits{D.}},
\bauthor{\bsnm{Thiele}, \binits{U.}}:
\batitle{Evolution of three-dimensional interface patterns in dewetting two-layer liquid films}.
\bjtitle{Europhys. Lett.}
\bvolume{74},
\bfpage{665}--\blpage{671}
(\byear{2006})
\doiurl{10.1209/epl/i2006-10026-8}
\end{barticle}
\endbibitem

\bibitem[\protect\citeauthoryear{Craster and Matar}{2009}]{CrMa2009rmp}
\begin{barticle}
\bauthor{\bsnm{Craster}, \binits{R.V.}},
\bauthor{\bsnm{Matar}, \binits{O.K.}}:
\batitle{Dynamics and stability of thin liquid films}.
\bjtitle{Rev. Mod. Phys.}
\bvolume{81},
\bfpage{1131}--\blpage{1198}
(\byear{2009})
\doiurl{10.1103/RevModPhys.81.1131}
\end{barticle}
\endbibitem

\bibitem[\protect\citeauthoryear{Ward}{2011}]{Ward2011pf}
\begin{barticle}
\bauthor{\bsnm{Ward}, \binits{M.H.}}:
\batitle{Interfacial thin films rupture and self-similarity}.
\bjtitle{Phys. Fluids}
\bvolume{23},
\bfpage{062105}
(\byear{2011})
\doiurl{10.1063/1.3604003}
\end{barticle}
\endbibitem

\bibitem[\protect\citeauthoryear{Bommer et~al.}{2013}]{BCJP2013epje}
\begin{barticle}
\bauthor{\bsnm{Bommer}, \binits{S.}},
\bauthor{\bsnm{Cartellier}, \binits{F.}},
\bauthor{\bsnm{Jachalski}, \binits{S.}},
\bauthor{\bsnm{Peschka}, \binits{D.}},
\bauthor{\bsnm{Seemann}, \binits{R.}},
\bauthor{\bsnm{Wagner}, \binits{B.}}:
\batitle{Droplets on liquids and their journey into equilibrium}.
\bjtitle{Eur. Phys. J. E}
\bvolume{36},
\bfpage{87}
(\byear{2013})
\doiurl{10.1140/epje/i2013-13087-x}
\end{barticle}
\endbibitem

\bibitem[\protect\citeauthoryear{Yadavali et~al.}{2012}]{YaKK2012prb}
\begin{barticle}
\bauthor{\bsnm{Yadavali}, \binits{S.}},
\bauthor{\bsnm{Krishna}, \binits{H.}},
\bauthor{\bsnm{Kalyanaraman}, \binits{R.}}:
\batitle{Morphology transitions in bilayer spinodal dewetting systems}.
\bjtitle{Phys. Rev. B}
\bvolume{85},
\bfpage{235446}
(\byear{2012})
\doiurl{10.1103/PhysRevB.85.235446}
\end{barticle}
\endbibitem

\bibitem[\protect\citeauthoryear{Jachalski et~al.}{2014}]{JPMW2014jem}
\begin{barticle}
\bauthor{\bsnm{Jachalski}, \binits{S.}},
\bauthor{\bsnm{Peschka}, \binits{D.}},
\bauthor{\bsnm{M{\"u}nch}, \binits{A.}},
\bauthor{\bsnm{Wagner}, \binits{B.}}:
\batitle{Impact of interfacial slip on the stability of liquid two-layer polymer films}.
\bjtitle{J. Eng. Math.}
\bvolume{86},
\bfpage{9}--\blpage{29}
(\byear{2014})
\doiurl{10.1007/s10665-013-9651-8}
\end{barticle}
\endbibitem

\bibitem[\protect\citeauthoryear{Xu et~al.}{2017}]{XBRS2017sr}
\begin{barticle}
\bauthor{\bsnm{Xu}, \binits{L.}},
\bauthor{\bsnm{Bandyopadhyay}, \binits{D.}},
\bauthor{\bsnm{Reddy}, \binits{P.D.S.}},
\bauthor{\bsnm{Sharma}, \binits{A.}},
\bauthor{\bsnm{Joo}, \binits{S.W.}}:
\batitle{Giant slip induced anomalous dewetting of an ultrathin film on a viscous sublayer}.
\bjtitle{Sci. Rep.}
\bvolume{7},
\bfpage{14776}
(\byear{2017})
\doiurl{10.1038/s41598-017-14861-4}
\end{barticle}
\endbibitem

\bibitem[\protect\citeauthoryear{Peschka et~al.}{2018}]{PBJS2018sr}
\begin{barticle}
\bauthor{\bsnm{Peschka}, \binits{D.}},
\bauthor{\bsnm{Bommer}, \binits{S.}},
\bauthor{\bsnm{Jachalski}, \binits{S.}},
\bauthor{\bsnm{Seemann}, \binits{R.}},
\bauthor{\bsnm{Wagner}, \binits{B.}}:
\batitle{Impact of energy dissipation on interface shapes and on rates for dewetting from liquid substrates}.
\bjtitle{Sci. Rep.}
\bvolume{8},
\bfpage{13295}
(\byear{2018})
\doiurl{10.1038/s41598-018-31418-1}
\end{barticle}
\endbibitem

\bibitem[\protect\citeauthoryear{L{\"u}dtge}{1869}]{Lued1869app}
\begin{barticle}
\bauthor{\bsnm{L{\"u}dtge}, \binits{R.}}:
\batitle{Ueber die {A}usbreitung der {F}l{\"u}ssigkeiten auf einander}.
\bjtitle{Ann. Phys. (Poggendorf)}
\bvolume{137},
\bfpage{362}--\blpage{377}
(\byear{1869})
\doiurl{10.1002/andp.18692130703}
\end{barticle}
\endbibitem

\bibitem[\protect\citeauthoryear{Craster and Matar}{2006}]{CrMa2006jcis}
\begin{barticle}
\bauthor{\bsnm{Craster}, \binits{R.V.}},
\bauthor{\bsnm{Matar}, \binits{O.K.}}:
\batitle{On the dynamics of liquid lenses}.
\bjtitle{J. Colloid Interface Sci.}
\bvolume{303},
\bfpage{503}--\blpage{516}
(\byear{2006})
\doiurl{10.1016/j.jcis.2006.08.009}
\end{barticle}
\endbibitem

\bibitem[\protect\citeauthoryear{Jachalski et~al.}{2013}]{JHKP2013sjam}
\begin{barticle}
\bauthor{\bsnm{Jachalski}, \binits{S.}},
\bauthor{\bsnm{Huth}, \binits{R.}},
\bauthor{\bsnm{Kitavtsev}, \binits{G.}},
\bauthor{\bsnm{Peschka}, \binits{D.}},
\bauthor{\bsnm{Wagner}, \binits{B.}}:
\batitle{Stationary solutions of liquid two-layer thin-film models}.
\bjtitle{SIAM J. Appl. Math.}
\bvolume{73},
\bfpage{1183}--\blpage{1202}
(\byear{2013})
\doiurl{10.1137/120886613}
\end{barticle}
\endbibitem

\bibitem[\protect\citeauthoryear{Huth et~al.}{2015}]{HJKP2015jem}
\begin{barticle}
\bauthor{\bsnm{Huth}, \binits{R.}},
\bauthor{\bsnm{Jachalski}, \binits{S.}},
\bauthor{\bsnm{Kitavtsev}, \binits{G.}},
\bauthor{\bsnm{Peschka}, \binits{D.}}:
\batitle{Gradient flow perspective on thin-film bilayer flows}.
\bjtitle{J. Eng. Math.}
\bvolume{94},
\bfpage{43}--\blpage{61}
(\byear{2015})
\doiurl{10.1007/s10665-014-9698-1}
\end{barticle}
\endbibitem

\bibitem[\protect\citeauthoryear{Kriegsmann and Miksis}{2003}]{KrMi2003sjam}
\begin{barticle}
\bauthor{\bsnm{Kriegsmann}, \binits{J.J.}},
\bauthor{\bsnm{Miksis}, \binits{M.J.}}:
\batitle{Steady motion of a drop along a liquid interface}.
\bjtitle{SIAM J. Appl. Math.}
\bvolume{64},
\bfpage{18}--\blpage{40}
(\byear{2003})
\doiurl{10.1137/S0036139901400215}
\end{barticle}
\endbibitem

\bibitem[\protect\citeauthoryear{Bertin et~al.}{2021}]{BLSR2021jfm}
\begin{barticle}
\bauthor{\bsnm{Bertin}, \binits{V.}},
\bauthor{\bsnm{Lee}, \binits{C.L.}},
\bauthor{\bsnm{Salez}, \binits{T.}},
\bauthor{\bsnm{Raphael}, \binits{E.}},
\bauthor{\bsnm{Dalnoki-Veress}, \binits{K.}}:
\batitle{Capillary levelling of immiscible bilayer films}.
\bjtitle{J. Fluid Mech.}
\bvolume{911},
\bfpage{13}
(\byear{2021})
\doiurl{10.1017/jfm.2020.1045}
\end{barticle}
\endbibitem

\bibitem[\protect\citeauthoryear{Mahadevan et~al.}{2002}]{MaAP2002jfm}
\begin{barticle}
\bauthor{\bsnm{Mahadevan}, \binits{L.}},
\bauthor{\bsnm{Adda-Bedia}, \binits{M.}},
\bauthor{\bsnm{Pomeau}, \binits{Y.}}:
\batitle{Four-phase merging in sessile compound drops}.
\bjtitle{J. Fluid Mech.}
\bvolume{451},
\bfpage{411}--\blpage{420}
(\byear{2002})
\doiurl{10.1017/S0022112001007108}
\end{barticle}
\endbibitem

\bibitem[\protect\citeauthoryear{Neeson et~al.}{2012}]{NTGD2012sm}
\begin{barticle}
\bauthor{\bsnm{Neeson}, \binits{M.J.}},
\bauthor{\bsnm{Tabor}, \binits{R.F.}},
\bauthor{\bsnm{Grieser}, \binits{F.}},
\bauthor{\bsnm{Dagastine}, \binits{R.R.}},
\bauthor{\bsnm{Chan}, \binits{D.Y.C.}}:
\batitle{Compound sessile drops}.
\bjtitle{Soft Matter}
\bvolume{8},
\bfpage{11042}--\blpage{11050}
(\byear{2012})
\doiurl{10.1039/c2sm26637g}
\end{barticle}
\endbibitem

\bibitem[\protect\citeauthoryear{Ben~Said et~al.}{2014}]{BSNB2014l}
\begin{barticle}
\bauthor{\bsnm{Ben~Said}, \binits{M.}},
\bauthor{\bsnm{Selzer}, \binits{M.}},
\bauthor{\bsnm{Nestler}, \binits{B.}},
\bauthor{\bsnm{Braun}, \binits{D.}},
\bauthor{\bsnm{Greiner}, \binits{C.}},
\bauthor{\bsnm{Garcke}, \binits{H.}}:
\batitle{A phase-field approach for wetting phenomena of multiphase droplets on solid surfaces}.
\bjtitle{Langmuir}
\bvolume{30},
\bfpage{4033}--\blpage{4039}
(\byear{2014})
\doiurl{10.1021/la500312q}
\end{barticle}
\endbibitem

\bibitem[\protect\citeauthoryear{Zhang et~al.}{2016}]{ZCAG2016jcis}
\begin{barticle}
\bauthor{\bsnm{Zhang}, \binits{Y.}},
\bauthor{\bsnm{Chatain}, \binits{D.}},
\bauthor{\bsnm{Anna}, \binits{S.L.}},
\bauthor{\bsnm{Garoff}, \binits{S.}}:
\batitle{Stability of a compound sessile drop at the axisymmetric configuration}.
\bjtitle{J. Colloid Interface Sci.}
\bvolume{462},
\bfpage{88}--\blpage{99}
(\byear{2016})
\doiurl{10.1016/j.jcis.2015.09.043}
\end{barticle}
\endbibitem

\bibitem[\protect\citeauthoryear{Iqbal et~al.}{2017}]{IDSS2017l}
\begin{barticle}
\bauthor{\bsnm{Iqbal}, \binits{R.}},
\bauthor{\bsnm{Dhiman}, \binits{S.}},
\bauthor{\bsnm{Sen}, \binits{A.K.}},
\bauthor{\bsnm{Shen}, \binits{A.Q.}}:
\batitle{Dynamics of a water droplet over a sessile oil droplet: {C}ompound droplets satisfying a {N}eumann condition}.
\bjtitle{Langmuir}
\bvolume{33},
\bfpage{5713}--\blpage{5723}
(\byear{2017})
\doiurl{10.1021/acs.langmuir.6b04621}
\end{barticle}
\endbibitem

\bibitem[\protect\citeauthoryear{Kitavtsev}{2024}]{Kita2024jns}
\begin{barticle}
\bauthor{\bsnm{Kitavtsev}, \binits{G.}}:
\batitle{Composite solutions to a liquid bilayer model}.
\bjtitle{J. Nonlinear Sci.}
\bvolume{35},
\bfpage{15}
(\byear{2024})
\doiurl{10.1007/s00332-024-10108-5}
\end{barticle}
\endbibitem

\bibitem[\protect\citeauthoryear{Nepomnyashchy}{2021}]{Nepo2021cocis}
\begin{barticle}
\bauthor{\bsnm{Nepomnyashchy}, \binits{A.}}:
\batitle{Droplet on a liquid substrate: {W}etting, dewetting, dynamics, instabilities}.
\bjtitle{Curr. Opin. Colloid Interface Sci.}
\bvolume{51},
\bfpage{101398}
(\byear{2021})
\doiurl{10.1016/j.cocis.2020.101398}
\end{barticle}
\endbibitem

\bibitem[\protect\citeauthoryear{Zhang et~al.}{2021}]{ZGLD2021jfm}
\begin{barticle}
\bauthor{\bsnm{Zhang}, \binits{C.Y.}},
\bauthor{\bsnm{Gao}, \binits{P.}},
\bauthor{\bsnm{Li}, \binits{E.Q.}},
\bauthor{\bsnm{Ding}, \binits{H.}}:
\batitle{On the compound sessile drops: {C}onfiguration boundaries and transitions}.
\bjtitle{J. Fluid Mech.}
\bvolume{917},
\bfpage{37}
(\byear{2021})
\doiurl{10.1017/jfm.2021.314}
\end{barticle}
\endbibitem

\bibitem[\protect\citeauthoryear{Yu et~al.}{2019}]{YKDL2019sm}
\begin{barticle}
\bauthor{\bsnm{Yu}, \binits{H.T.}},
\bauthor{\bsnm{Kant}, \binits{P.}},
\bauthor{\bsnm{Dyett}, \binits{B.}},
\bauthor{\bsnm{Lohse}, \binits{D.}},
\bauthor{\bsnm{Zhang}, \binits{X.H.}}:
\batitle{Splitting droplets through coalescence of two different three-phase contact lines}.
\bjtitle{Soft Matter}
\bvolume{15},
\bfpage{6055}--\blpage{6061}
(\byear{2019})
\doiurl{10.1039/c9sm00638a}
\end{barticle}
\endbibitem

\bibitem[\protect\citeauthoryear{Danov et~al.}{1998a}]{DPAR1998ces}
\begin{barticle}
\bauthor{\bsnm{Danov}, \binits{K.D.}},
\bauthor{\bsnm{Paunov}, \binits{V.N.}},
\bauthor{\bsnm{Alleborn}, \binits{N.}},
\bauthor{\bsnm{Raszillier}, \binits{H.}},
\bauthor{\bsnm{Durst}, \binits{F.}}:
\batitle{Stability of evaporating two-layered liquid film in the presence of surfactant - {I. T}he equations of lubrication approximation}.
\bjtitle{Chem. Eng. Sci.}
\bvolume{53},
\bfpage{2809}--\blpage{2822}
(\byear{1998})
\doiurl{10.1016/s0009-2509(98)00098-0}
\end{barticle}
\endbibitem

\bibitem[\protect\citeauthoryear{Danov et~al.}{1998b}]{DPSA1998ces}
\begin{barticle}
\bauthor{\bsnm{Danov}, \binits{K.D.}},
\bauthor{\bsnm{Paunov}, \binits{V.N.}},
\bauthor{\bsnm{Stoyanov}, \binits{S.D.}},
\bauthor{\bsnm{Alleborn}, \binits{N.}},
\bauthor{\bsnm{Raszillier}, \binits{H.}},
\bauthor{\bsnm{Durst}, \binits{F.}}:
\batitle{Stability of evaporating two-layered liquid film in the presence of surfactant - {II. L}inear analysis}.
\bjtitle{Chem. Eng. Sci.}
\bvolume{53},
\bfpage{2823}--\blpage{2837}
(\byear{1998})
\doiurl{10.1016/s0009-2509(98)00099-2}
\end{barticle}
\endbibitem

\bibitem[\protect\citeauthoryear{Paunov et~al.}{1998}]{PDAR1998ces}
\begin{barticle}
\bauthor{\bsnm{Paunov}, \binits{V.N.}},
\bauthor{\bsnm{Danov}, \binits{K.D.}},
\bauthor{\bsnm{Alleborn}, \binits{N.}},
\bauthor{\bsnm{Raszillier}, \binits{H.}},
\bauthor{\bsnm{Durst}, \binits{F.}}:
\batitle{Stability of evaporating two-layered liquid film in the presence of surfactant - {III. N}on-linear stability analysis}.
\bjtitle{Chem. Eng. Sci.}
\bvolume{53},
\bfpage{2839}--\blpage{2857}
(\byear{1998})
\doiurl{10.1016/S0009-2509(98)00100-6}
\end{barticle}
\endbibitem

\bibitem[\protect\citeauthoryear{Mukherjee and Sharma}{2015}]{MuSh2015sm}
\begin{barticle}
\bauthor{\bsnm{Mukherjee}, \binits{R.}},
\bauthor{\bsnm{Sharma}, \binits{A.}}:
\batitle{Instability, self-organization and pattern formation in thin soft films}.
\bjtitle{Soft Matter}
\bvolume{11},
\bfpage{8717}--\blpage{8740}
(\byear{2015})
\doiurl{10.1039/c5sm01724f}
\end{barticle}
\endbibitem

\bibitem[\protect\citeauthoryear{Nepomnyashchy and Simanovskii}{2009}]{NeSi2009prl}
\begin{barticle}
\bauthor{\bsnm{Nepomnyashchy}, \binits{A.}},
\bauthor{\bsnm{Simanovskii}, \binits{I.}}:
\batitle{Instabilities and ordered patterns in nonisothermal ultrathin bilayer fluid films}.
\bjtitle{Phys. Rev. Lett.}
\bvolume{102},
\bfpage{164501}
(\byear{2009})
\doiurl{10.1103/PhysRevLett.102.164501}
\end{barticle}
\endbibitem

\bibitem[\protect\citeauthoryear{Nepomnyashchy and Simanovskii}{2017}]{NeSi2017pf}
\begin{barticle}
\bauthor{\bsnm{Nepomnyashchy}, \binits{A.A.}},
\bauthor{\bsnm{Simanovskii}, \binits{I.B.}}:
\batitle{Novel criteria for the development of monotonic and oscillatory instabilities in a two-layer film}.
\bjtitle{Phys. Fluids}
\bvolume{29},
\bfpage{092104}
(\byear{2017})
\doiurl{10.1063/1.5001729}
\end{barticle}
\endbibitem

\bibitem[\protect\citeauthoryear{Nepomnyashchy and Simanovskii}{2021}]{NeSi2021prf}
\begin{barticle}
\bauthor{\bsnm{Nepomnyashchy}, \binits{A.}},
\bauthor{\bsnm{Simanovskii}, \binits{I.}}:
\batitle{Droplets on the liquid substrate: {T}hermocapillary oscillatory instability}.
\bjtitle{Phys. Rev. Fluids}
\bvolume{6},
\bfpage{034001}
(\byear{2021})
\doiurl{10.1103/PhysRevFluids.6.034001}
\end{barticle}
\endbibitem

\bibitem[\protect\citeauthoryear{Geoghegan and Krausch}{2003}]{GeKr2003pps}
\begin{barticle}
\bauthor{\bsnm{Geoghegan}, \binits{M.}},
\bauthor{\bsnm{Krausch}, \binits{G.}}:
\batitle{Wetting at polymer surfaces and interfaces}.
\bjtitle{Prog. Polym. Sci.}
\bvolume{28},
\bfpage{261}--\blpage{302}
(\byear{2003})
\doiurl{10.1016/S0079-6700(02)00080-1}
\end{barticle}
\endbibitem

\bibitem[\protect\citeauthoryear{Govor et~al.}{2006}]{GRBP2006pla}
\begin{barticle}
\bauthor{\bsnm{Govor}, \binits{L.V.}},
\bauthor{\bsnm{Reiter}, \binits{G.}},
\bauthor{\bsnm{Bauer}, \binits{G.H.}},
\bauthor{\bsnm{Parisi}, \binits{J.}}:
\batitle{Bilayer formation in thin films of a binary solution}.
\bjtitle{Phys. Lett. A}
\bvolume{353},
\bfpage{198}--\blpage{204}
(\byear{2006})
\doiurl{10.1016/j.physleta.2005.11.088}
\end{barticle}
\endbibitem

\bibitem[\protect\citeauthoryear{N{\'a}raigh and Thiffeault}{2010}]{NaTh2010n}
\begin{barticle}
\bauthor{\bsnm{N{\'a}raigh}, \binits{L.{\'O}.}},
\bauthor{\bsnm{Thiffeault}, \binits{J.L.}}:
\batitle{Nonlinear dynamics of phase separation in thin films}.
\bjtitle{Nonlinearity}
\bvolume{23},
\bfpage{1559}--\blpage{1583}
(\byear{2010})
\doiurl{10.1088/0951-7715/23/7/003}
\end{barticle}
\endbibitem

\bibitem[\protect\citeauthoryear{Thiele et~al.}{2013}]{ThTL2013prl}
\begin{barticle}
\bauthor{\bsnm{Thiele}, \binits{U.}},
\bauthor{\bsnm{Todorova}, \binits{D.V.}},
\bauthor{\bsnm{Lopez}, \binits{H.}}:
\batitle{Gradient dynamics description for films of mixtures and suspensions: dewetting triggered by coupled film height and concentration fluctuations}.
\bjtitle{Phys. Rev. Lett.}
\bvolume{111},
\bfpage{117801}
(\byear{2013})
\doiurl{10.1103/PhysRevLett.111.117801}
\end{barticle}
\endbibitem

\bibitem[\protect\citeauthoryear{Diez et~al.}{2021}]{DGGR2021l}
\begin{barticle}
\bauthor{\bsnm{Diez}, \binits{J.A.}},
\bauthor{\bsnm{González}, \binits{A.G.}},
\bauthor{\bsnm{Garfinkel}, \binits{D.A.}},
\bauthor{\bsnm{Rack}, \binits{P.D.}},
\bauthor{\bsnm{McKeown}, \binits{J.T.}},
\bauthor{\bsnm{Kondic}, \binits{L.}}:
\batitle{Simultaneous decomposition and dewetting of nanoscale alloys: {A} comparison of experiment and theory}.
\bjtitle{Langmuir}
\bvolume{37}(\bissue{8}),
\bfpage{2575}--\blpage{2585}
(\byear{2021})
\doiurl{10.1021/acs.langmuir.0c02964}
\end{barticle}
\endbibitem

\bibitem[\protect\citeauthoryear{Areshi et~al.}{2024}]{ATTG2024pre}
\begin{barticle}
\bauthor{\bsnm{Areshi}, \binits{M.}},
\bauthor{\bsnm{Tseluiko}, \binits{D.}},
\bauthor{\bsnm{Thiele}, \binits{U.}},
\bauthor{\bsnm{Goddard}, \binits{B.D.}},
\bauthor{\bsnm{Archer}, \binits{A.J.}}:
\batitle{Binding potential and wetting behavior of binary liquid mixtures on surfaces}.
\bjtitle{Phys. Rev. E}
\bvolume{109},
\bfpage{024801}
(\byear{2024})
\doiurl{10.1103/PhysRevE.109.024801}
\end{barticle}
\endbibitem

\bibitem[\protect\citeauthoryear{Oron et~al.}{1997}]{OrDB1997rmp}
\begin{barticle}
\bauthor{\bsnm{Oron}, \binits{A.}},
\bauthor{\bsnm{Davis}, \binits{S.H.}},
\bauthor{\bsnm{Bankoff}, \binits{S.G.}}:
\batitle{Long-scale evolution of thin liquid films}.
\bjtitle{Rev. Mod. Phys.}
\bvolume{69},
\bfpage{931}--\blpage{980}
(\byear{1997})
\doiurl{10.1103/RevModPhys.69.931}
\end{barticle}
\endbibitem

\bibitem[\protect\citeauthoryear{Benney}{1966}]{Benn1966jmp}
\begin{barticle}
\bauthor{\bsnm{Benney}, \binits{D.J.}}:
\batitle{Long waves on liquid films}.
\bjtitle{J. Math. Phys.}
\bvolume{45},
\bfpage{150}--\blpage{155}
(\byear{1966})
\doiurl{10.1002/sapm1966451150}
\end{barticle}
\endbibitem

\bibitem[\protect\citeauthoryear{Scheid et~al.}{2005}]{SRTK2005jfm}
\begin{barticle}
\bauthor{\bsnm{Scheid}, \binits{B.}},
\bauthor{\bsnm{Ruyer-Quil}, \binits{C.}},
\bauthor{\bsnm{Thiele}, \binits{U.}},
\bauthor{\bsnm{Kabov}, \binits{O.A.}},
\bauthor{\bsnm{Legros}, \binits{J.C.}},
\bauthor{\bsnm{Colinet}, \binits{P.}}:
\batitle{Validity domain of the {B}enney equation including {M}arangoni effect for closed and open flows}.
\bjtitle{J. Fluid Mech.}
\bvolume{527},
\bfpage{303}--\blpage{335}
(\byear{2005})
\doiurl{10.1017/S0022112004003179}
\end{barticle}
\endbibitem

\bibitem[\protect\citeauthoryear{Çekiç and Sisoev}{2015}]{CeSi2015ijonm}
\begin{barticle}
\bauthor{\bsnm{Çekiç}, \binits{G.}},
\bauthor{\bsnm{Sisoev}, \binits{G.M.}}:
\batitle{On wave families in a two-layer falling film}.
\bjtitle{International Journal of Non-Linear Mechanics}
\bvolume{69},
\bfpage{45}--\blpage{54}
(\byear{2015})
\doiurl{10.1016/j.ijnonlinmec.2014.11.020}
\end{barticle}
\endbibitem

\bibitem[\protect\citeauthoryear{Kliakhandler}{1999}]{Klia1999jfm}
\begin{barticle}
\bauthor{\bsnm{Kliakhandler}, \binits{I.L.}}:
\batitle{Long interfacial waves in multilayer thin films and coupled {K}uramoto-{S}ivashinsky equations}.
\bjtitle{J. Fluid Mech.}
\bvolume{391},
\bfpage{45}--\blpage{65}
(\byear{1999})
\doiurl{10.1017/S0022112099005297}
\end{barticle}
\endbibitem

\bibitem[\protect\citeauthoryear{Yu}{2026}]{Yu2026wm}
\begin{barticle}
\bauthor{\bsnm{Yu}, \binits{B.}}:
\batitle{Two-layer roll waves in rapid gravity currents on mild slopes}.
\bjtitle{Wave Motion}
\bvolume{140},
\bfpage{103634}
(\byear{2026})
\doiurl{10.1016/j.wavemoti.2025.103634}
\end{barticle}
\endbibitem

\bibitem[\protect\citeauthoryear{Pototsky and Maksymov}{2023}]{PoMa2023prf}
\begin{barticle}
\bauthor{\bsnm{Pototsky}, \binits{A.}},
\bauthor{\bsnm{Maksymov}, \binits{I.S.}}:
\batitle{Nonlinear periodic and solitary rolling waves in falling two-layer viscous liquid films}.
\bjtitle{Phys. Rev. Fluids}
\bvolume{8},
\bfpage{064801}
(\byear{2023})
\doiurl{10.1103/PhysRevFluids.8.064801}
\end{barticle}
\endbibitem

\bibitem[\protect\citeauthoryear{Neumann and Wangerin}{1894}]{NeumannWangerin1894}
\begin{bbook}
\bauthor{\bsnm{Neumann}, \binits{F.E.}},
\bauthor{\bsnm{Wangerin}, \binits{A.}}:
\bbtitle{Vorlesungen {\"U}ber die {T}heorie der {C}apillarit{\"a}t, Gehalten an der {U}niversit{\"a}t {K}{\"o}nigsberg von {F}ranz {N}eumann}.
\bsertitle{Vorlesungen {\"u}ber mathematische {P}hysik. (7. {H}ft.)}.
\bpublisher{B. G. Teubner},
\blocation{Leipzig}
(\byear{1894})
\end{bbook}
\endbibitem

\bibitem[\protect\citeauthoryear{Diekmann and Thiele}{2025}]{DiTh2025prf}
\begin{barticle}
\bauthor{\bsnm{Diekmann}, \binits{J.}},
\bauthor{\bsnm{Thiele}, \binits{U.}}:
\batitle{Mesoscopic hydrodynamic model for spreading, sliding and coarsening compound drops}.
\bjtitle{Phys. Rev. Fluids}
\bvolume{10},
\bfpage{024002}
(\byear{2025})
\doiurl{10.1103/PhysRevFluids.10.024002}
\end{barticle}
\endbibitem

\bibitem[\protect\citeauthoryear{Thiele}{2018}]{Thie2018csa}
\begin{barticle}
\bauthor{\bsnm{Thiele}, \binits{U.}}:
\batitle{Recent advances in and future challenges for mesoscopic hydrodynamic modelling of complex wetting}.
\bjtitle{Colloid Surf. A}
\bvolume{553},
\bfpage{487}--\blpage{495}
(\byear{2018})
\doiurl{10.1016/j.colsurfa.2018.05.049}
\end{barticle}
\endbibitem

\bibitem[\protect\citeauthoryear{Heil and Hazel}{2006}]{HeHa2006}
\begin{bchapter}
\bauthor{\bsnm{Heil}, \binits{M.}},
\bauthor{\bsnm{Hazel}, \binits{A.L.}}:
\bctitle{Oomph-lib - an object-oriented multi-physics finite-element library}.
In: \beditor{\bsnm{Bungartz}, \binits{H.-J.}},
\beditor{\bsnm{Sch{\"a}fer}, \binits{M.}} (eds.)
\bbtitle{Fluid-Structure Interaction: Modelling, Simulation, Optimisation},
pp. \bfpage{19}--\blpage{49}.
\bpublisher{Springer},
\blocation{Berlin, Heidelberg}
(\byear{2006}).
\doiurl{10.1007/3-540-34596-5_2}
\end{bchapter}
\endbibitem

\bibitem[\protect\citeauthoryear{Uecker et~al.}{2014}]{UeWR2014nmma}
\begin{barticle}
\bauthor{\bsnm{Uecker}, \binits{H.}},
\bauthor{\bsnm{Wetzel}, \binits{D.}},
\bauthor{\bsnm{Rademacher}, \binits{J.D.M.}}:
\batitle{{pde2path} - a {Matlab} package for continuation and bifurcation in {2D} elliptic systems}.
\bjtitle{Numer. Math.-Theory Methods Appl.}
\bvolume{7},
\bfpage{58}--\blpage{106}
(\byear{2014})
\doiurl{10.4208/nmtma.2014.1231nm}
\end{barticle}
\endbibitem

\bibitem[\protect\citeauthoryear{Engelnkemper et~al.}{2019}]{EGUW2019springer}
\begin{bchapter}
\bauthor{\bsnm{Engelnkemper}, \binits{S.}},
\bauthor{\bsnm{Gurevich}, \binits{S.V.}},
\bauthor{\bsnm{Uecker}, \binits{H.}},
\bauthor{\bsnm{Wetzel}, \binits{D.}},
\bauthor{\bsnm{Thiele}, \binits{U.}}:
\bctitle{Continuation for thin film hydrodynamics and related scalar problems}.
In: \beditor{\bsnm{Gelfgat}, \binits{A.}} (ed.)
\bbtitle{Computational Modeling of Bifurcations and Instabilities in Fluid Mechanics}.
\bsertitle{Computational Methods in Applied Sciences, vol 50},
pp. \bfpage{459}--\blpage{501}.
\bpublisher{Springer},
\blocation{Cham}
(\byear{2019}).
\doiurl{10.1007/978-3-319-91494-7_13}
\end{bchapter}
\endbibitem

\bibitem[\protect\citeauthoryear{Diddens and Rocha}{2024}]{DiRo2024jocp}
\begin{barticle}
\bauthor{\bsnm{Diddens}, \binits{C.}},
\bauthor{\bsnm{Rocha}, \binits{D.}}:
\batitle{Bifurcation tracking on moving meshes and with consideration of azimuthal symmetry breaking instabilities}.
\bjtitle{Journal of Computational Physics}
\bvolume{518},
\bfpage{113306}
(\byear{2024})
\doiurl{10.1016/j.jcp.2024.113306}
\end{barticle}
\endbibitem

\bibitem[\protect\citeauthoryear{Israelachvili}{2011}]{Israelachvili2011}
\begin{bbook}
\bauthor{\bsnm{Israelachvili}, \binits{J.N.}}:
\bbtitle{Intermolecular and Surface Forces},
\bedition{3rd} edn.
\bpublisher{Academic Press},
\blocation{San Diego}
(\byear{2011})
\end{bbook}
\endbibitem

\bibitem[\protect\citeauthoryear{Bauer et~al.}{2002}]{BaFK2002josc}
\begin{barticle}
\bauthor{\bsnm{Bauer}, \binits{C.}},
\bauthor{\bsnm{Frink}, \binits{A.}},
\bauthor{\bsnm{Kreckel}, \binits{R.}}:
\batitle{Introduction to the {GiNaC} framework for symbolic computation within the {C}++ programming language}.
\bjtitle{J. Symb. Comput.}
\bvolume{33},
\bfpage{1}--\blpage{12}
(\byear{2002})
\doiurl{10.1006/jsco.2001.0494}
\end{barticle}
\endbibitem

\bibitem[\protect\citeauthoryear{Henkel et~al.}{2021}]{HeST2021sm}
\begin{barticle}
\bauthor{\bsnm{Henkel}, \binits{C.}},
\bauthor{\bsnm{Snoeijer}, \binits{J.H.}},
\bauthor{\bsnm{Thiele}, \binits{U.}}:
\batitle{Gradient-dynamics model for liquid drops on elastic substrates}.
\bjtitle{Soft Matter}
\bvolume{17},
\bfpage{10359}--\blpage{10375}
(\byear{2021})
\doiurl{10.1039/D1SM01032H}
\end{barticle}
\endbibitem

\bibitem[\protect\citeauthoryear{Thy}{2024}]{BT_Thy_2024}
\begin{botherref}
\oauthor{\bsnm{Thy}, \binits{D.}}:
Modeling of interface-dominated sliding compound drops.
Bachelor's thesis,
University of Münster
(2024)
\end{botherref}
\endbibitem

\bibitem[\protect\citeauthoryear{Moradi et~al.}{2011}]{MoVS2011el}
\begin{barticle}
\bauthor{\bsnm{Moradi}, \binits{N.}},
\bauthor{\bsnm{Varnik}, \binits{F.}},
\bauthor{\bsnm{Steinbach}, \binits{I.}}:
\batitle{Contact angle dependence of the velocity of sliding cylindrical drop on flat substrates}.
\bjtitle{Europhys. Lett.}
\bvolume{95},
\bfpage{44003}
(\byear{2011})
\doiurl{10.1209/0295-5075/95/44003}
\end{barticle}
\endbibitem

\bibitem[\protect\citeauthoryear{Strogatz}{2015}]{Strogatz2015}
\begin{bbook}
\bauthor{\bsnm{Strogatz}, \binits{S.H.}}:
\bbtitle{Nonlinear Dynamics and Chaos},
\bedition{2nd} edn.
\bpublisher{CRC Press},
\blocation{Boca Raton}
(\byear{2015}).
\doiurl{10.1201/9780429492563}
\end{bbook}
\endbibitem

\bibitem[\protect\citeauthoryear{Tuckerman and Barkley}{1988}]{TuBa1988prl}
\begin{barticle}
\bauthor{\bsnm{Tuckerman}, \binits{L.S.}},
\bauthor{\bsnm{Barkley}, \binits{D.}}:
\batitle{Global bifurcation to traveling waves in axisymmetric convection}.
\bjtitle{Phys. Rev. Lett.}
\bvolume{61},
\bfpage{408}--\blpage{411}
(\byear{1988})
\doiurl{10.1103/PhysRevLett.61.408}
\end{barticle}
\endbibitem

\bibitem[\protect\citeauthoryear{Thiele and Knobloch}{2006}]{ThKn2006njp}
\begin{barticle}
\bauthor{\bsnm{Thiele}, \binits{U.}},
\bauthor{\bsnm{Knobloch}, \binits{E.}}:
\batitle{On the depinning of a driven drop on a heterogeneous substrate}.
\bjtitle{New J. Phys.}
\bvolume{8},
\bfpage{313}
(\byear{2006})
\doiurl{10.1088/1367-2630/8/12/313}
\end{barticle}
\endbibitem

\bibitem[\protect\citeauthoryear{Marques et~al.}{2013}]{MMLP2013prsapes}
\begin{barticle}
\bauthor{\bsnm{Marques}, \binits{F.}},
\bauthor{\bsnm{Meseguer}, \binits{A.}},
\bauthor{\bsnm{Lopez}, \binits{J.M.}},
\bauthor{\bsnm{Pacheco}, \binits{J.R.}},
\bauthor{\bsnm{Lopez}, \binits{J.M.}}:
\batitle{Bifurcations with imperfect {SO}(2) symmetry and pinning of rotating waves}.
\bjtitle{Proc. R. Soc. A-Math. Phys. Eng. Sci.}
\bvolume{469},
\bfpage{20120348}
(\byear{2013})
\doiurl{10.1098/rspa.2012.0348}
\end{barticle}
\endbibitem

\bibitem[\protect\citeauthoryear{K{\"o}pf and Thiele}{2014}]{KoTh2014n}
\begin{barticle}
\bauthor{\bsnm{K{\"o}pf}, \binits{M.H.}},
\bauthor{\bsnm{Thiele}, \binits{U.}}:
\batitle{Emergence of the bifurcation structure of a {L}angmuir-{B}lodgett transfer model}.
\bjtitle{Nonlinearity}
\bvolume{27},
\bfpage{2711}--\blpage{2734}
(\byear{2014})
\doiurl{10.1088/0951-7715/27/11/2711}
{\href{https://arxiv.org/abs/http://arxiv.org/abs/1405.2117}{{http://arxiv.org/abs/1405.2117}}}
\end{barticle}
\endbibitem

\bibitem[\protect\citeauthoryear{Grawitter and Stark}{2024}]{GrSt2024sm}
\begin{barticle}
\bauthor{\bsnm{Grawitter}, \binits{J.}},
\bauthor{\bsnm{Stark}, \binits{H.}}:
\batitle{Steering droplets on substrates with plane-wave wettability patterns and deformations}.
\bjtitle{Soft Matter}
\bvolume{20},
\bfpage{3161}--\blpage{3174}
(\byear{2024})
\doiurl{10.1039/d4sm00213j}
\end{barticle}
\endbibitem

\bibitem[\protect\citeauthoryear{Lubbers et~al.}{2014}]{LWBD2014jfm}
\begin{barticle}
\bauthor{\bsnm{Lubbers}, \binits{L.A.}},
\bauthor{\bsnm{Weijs}, \binits{J.H.}},
\bauthor{\bsnm{Botto}, \binits{L.}},
\bauthor{\bsnm{Das}, \binits{S.}},
\bauthor{\bsnm{Andreotti}, \binits{B.}},
\bauthor{\bsnm{Snoeijer}, \binits{J.H.}}:
\batitle{Drops on soft solids: free energy and double transition of contact angles}.
\bjtitle{J. Fluid Mech.}
\bvolume{747},
\bfpage{1}
(\byear{2014})
\doiurl{10.1017/jfm.2014.152}
\end{barticle}
\endbibitem

\bibitem[\protect\citeauthoryear{Greve et~al.}{2023}]{GrHT2023sm}
\begin{barticle}
\bauthor{\bsnm{Greve}, \binits{D.}},
\bauthor{\bsnm{Hartmann}, \binits{S.}},
\bauthor{\bsnm{Thiele}, \binits{U.}}:
\batitle{Stick-slip dynamics in the forced wetting of polymer brushes}.
\bjtitle{Soft Matter}
\bvolume{19},
\bfpage{4041}--\blpage{4061}
(\byear{2023})
\doiurl{10.1039/D3SM00104K}
\end{barticle}
\endbibitem

\bibitem[\protect\citeauthoryear{Lin et~al.}{2001}]{LKBH2001jcp}
\begin{barticle}
\bauthor{\bsnm{Lin}, \binits{Z.}},
\bauthor{\bsnm{Kerle}, \binits{T.}},
\bauthor{\bsnm{Baker}, \binits{S.M.}},
\bauthor{\bsnm{Hoagland}, \binits{D.A.}},
\bauthor{\bsnm{Sch{\"a}ffer}, \binits{E.}},
\bauthor{\bsnm{Steiner}, \binits{U.}},
\bauthor{\bsnm{Russell}, \binits{T.P.}}:
\batitle{Electric field induced instabilities at liquid/liquid interfaces}.
\bjtitle{J. Chem. Phys.}
\bvolume{114},
\bfpage{2377}--\blpage{2381}
(\byear{2001})
\doiurl{10.1063/1.1338125}
\end{barticle}
\endbibitem

\bibitem[\protect\citeauthoryear{Verma et~al.}{2005}]{VSKB2005l}
\begin{barticle}
\bauthor{\bsnm{Verma}, \binits{R.}},
\bauthor{\bsnm{Sharma}, \binits{A.}},
\bauthor{\bsnm{Kargupta}, \binits{K.}},
\bauthor{\bsnm{Bhaumik}, \binits{J.}}:
\batitle{Electric field induced instability and pattern formation in thin liquid films}.
\bjtitle{Langmuir}
\bvolume{21},
\bfpage{3710}--\blpage{3721}
(\byear{2005})
\doiurl{10.1021/la0472100}
\end{barticle}
\endbibitem

\bibitem[\protect\citeauthoryear{Amarandei et~al.}{2012}]{ABCO2012sm}
\begin{barticle}
\bauthor{\bsnm{Amarandei}, \binits{G.}},
\bauthor{\bsnm{Beltrame}, \binits{P.}},
\bauthor{\bsnm{Clancy}, \binits{I.}},
\bauthor{\bsnm{O'Dwyer}, \binits{C.}},
\bauthor{\bsnm{Arshak}, \binits{A.}},
\bauthor{\bsnm{Steiner}, \binits{U.}},
\bauthor{\bsnm{Corcoran}, \binits{D.}},
\bauthor{\bsnm{Thiele}, \binits{U.}}:
\batitle{Pattern formation induced by an electric field in a polymer-air-polymer thin film system}.
\bjtitle{Soft Matter}
\bvolume{8},
\bfpage{6333}--\blpage{6349}
(\byear{2012})
\doiurl{10.1039/c2sm25273b}
\end{barticle}
\endbibitem

\bibitem[\protect\citeauthoryear{Matar and Craster}{2009}]{MaCr2009sm}
\begin{barticle}
\bauthor{\bsnm{Matar}, \binits{O.K.}},
\bauthor{\bsnm{Craster}, \binits{R.V.}}:
\batitle{Dynamics of surfactant-assisted spreading}.
\bjtitle{Soft Matter}
\bvolume{5},
\bfpage{3801}--\blpage{3809}
(\byear{2009})
\doiurl{10.1039/b908719m}
\end{barticle}
\endbibitem

\bibitem[\protect\citeauthoryear{Thiele et~al.}{2016}]{ThAP2016prf}
\begin{barticle}
\bauthor{\bsnm{Thiele}, \binits{U.}},
\bauthor{\bsnm{Archer}, \binits{A.J.}},
\bauthor{\bsnm{Pismen}, \binits{L.M.}}:
\batitle{Gradient dynamics models for liquid films with soluble surfactant}.
\bjtitle{Phys. Rev. Fluids}
\bvolume{1},
\bfpage{083903}
(\byear{2016})
\doiurl{10.1103/PhysRevFluids.1.083903}
\end{barticle}
\endbibitem

\bibitem[\protect\citeauthoryear{Thiele}{2007}]{Thie2007chapter}
\begin{bchapter}
\bauthor{\bsnm{Thiele}, \binits{U.}}:
\bctitle{Structure formation in thin liquid films}.
In: \beditor{\bsnm{Kalliadasis}, \binits{S.}},
\beditor{\bsnm{Thiele}, \binits{U.}} (eds.)
\bbtitle{Thin Films of Soft Matter},
pp. \bfpage{25}--\blpage{93}.
\bpublisher{Springer},
\blocation{Vienna}
(\byear{2007}).
\doiurl{10.1007/978-3-211-69808-2\_2}
\end{bchapter}
\endbibitem

\end{thebibliography}
\end{document}